\documentclass[10pt,aps,prd,twocolumn, nofootinbib]{revtex4-1}

\usepackage{amssymb,amsmath}
\usepackage{physics}
\usepackage{subfig}
\usepackage{graphicx}
\usepackage{color}
\usepackage{hyperref}
\usepackage{listings}
\usepackage{braket} 
\usepackage{tablefootnote}
\hypersetup{
    bookmarks=true,         % show bookmarks bar?
    unicode=false,          % non-Latin characters in Acrobat’s bookmarks
    pdftoolbar=true,        % show Acrobat’s toolbar?
    pdfmenubar=true,        % show Acrobat’s menu?
    pdffitwindow=false,     % window fit to page when opened
    pdfstartview={FitH},    % fits the width of the page to the window
    pdfauthor={Kevin Multani},     % author
    colorlinks=true,       % false: boxed links; true: colored links
    linkcolor=blue,          % color of internal links
    citecolor=red,        % color of links to bibliography
}
\graphicspath{{figures/}}

\newcommand*\dif{\mathop{}\!\mathrm{d}}
\begin{document}

\definecolor{dkgreen}{rgb}{0,0.6,0}
\definecolor{gray}{rgb}{0.5,0.5,0.5}
\definecolor{mauve}{rgb}{0.58,0,0.82}

\lstset{frame=tb,
  	language=Matlab,
  	aboveskip=3mm,
  	belowskip=3mm,
  	showstringspaces=false,
  	columns=flexible,
  	basicstyle={\small\ttfamily},
  	numbers=none,
  	numberstyle=\tiny\color{gray},
 	keywordstyle=\color{blue},
	commentstyle=\color{dkgreen},
  	stringstyle=\color{mauve},
  	breaklines=true,
  	breakatwhitespace=true
  	tabsize=3
}

\title{Chiral-vacuum excited replicae in QCD modeling}

\author{Eduardo Garnacho-Velasco}
\affiliation{Universität Bielefeld,
Universit\"atsstra{\ss}e 25, 33615 Bielefeld, Germany}

\author{Pedro J. de A. Bicudo and J. Emilio F. T. Ribeiro}
\affiliation{CeFEMA, Departamento de F\'{\i}sica, Instituto Superior T\'ecnico, Universidade de Lisboa, Portugal}

\author{Felipe J. Llanes-Estrada, Lucas P\'erez Molina$^*$, Victor Serrano Herreros$^\dagger$, Jorge Vallejo Fern\'andez$^\&$}
\affiliation{Dept. F\'{\i}sica Te\'orica \& IPARCOS, Universidad Complutense de Madrid, Spain}
\altaffiliation{
Newly at Univ. de Barcelona, Facultad de F\'{\i}sica;
$^\dagger$ Newly at Rhein Westf\"alische Technische Hochschule Aachen, department of physics; $^\&$ Newly at Univ. Rey Juan Carlos. }

\date{\today}

\begin{abstract}
We present a detailed study of the Bardeen-Cooper-Schrieffer (BCS) gap equation ``replicae'' or excited vacuum states, orthogonal to the ground-state one, in the chiral-quark sector of the Hamiltonian Coulomb-gauge model of chromodynamics. Analyzing the number of negative eigenmodes of the energy density's Hessian we believe that we have identified all of the (negative energy-density) vacua of this nonlinear system, namely the ground BCS state and two (or one) replicae for slightly massive (or massless) quarks, given the interaction strength typical of the strong interactions. The meson spectrum over each of the replicae looks similar, so the differences are not significant enough given model uncertainties, but matrix elements are more sensitive and allow to distinguish them. We propose to look for such excited vacua in lattice gauge theory by trying to identify excitations with scalar quantum numbers
which have energies proportional to the lattice volume (unlike conventional mesons for which the mass stabilizes to a constant upon taking the infinite volume limit). 
\end{abstract}

\maketitle
%%%%%%%%%%%%%%%%%%%%%%%%%%%%%%%%%%%%%%%%%%%%%%%%%%%%%%%%%%%%%%
\section{Introduction}
%%%%%%%%%%%%%%%%%%%%%%%%%%%%%%%%%%%%%%%%%%%%%%%%%%%%%%%%%%%%%%

False vacua are an intriguing concept. The possibility that a tunnelling event  may suddenly  ``open the ground under our feet''
or more practically, that such sudden phase transitions may have happened in the history of the universe, and certainly happen in condensed matter systems, is of great interest.

In the strong interactions, spontaneous chiral symmetry breaking happens as the ground state, driven in equal--time quantization by a scalar quark-antiquark condensate, is not annihilated by the chiral charge. The gap equation~\cite{Adler:1984ri,Bicudo:1989sh}  that describes that condensate and sets the vacuum state of the interaction theory in terms of the degrees of freedom of the noninteracting Hamiltonian, was discovered to have more than one solution two decades ago~\cite{Bicudo:2002eu}. The phenomenon was also reported for fermion-scalar Dyson-Schwinger integral equations with strong enough couplings  ~\cite{Llanes-Estrada:2006bxu}.

The spectrum of single mesons built over those ``replicae'' of the vacuum in the chiral symmetry breaking sector of the Standard Model
has been found to be definite positive~\cite{Bicudo:2019ryc}, so that few-body excitations over those vacua appear as conventional hadron spectra. In this article we show that \emph{collective} modes include a negative eigenvalue and thus lead to vacuum instability that
takes the excited vacuum to the ground state by a continuous, monotonous, energy-decreasing  trajectory. The number and nature of relevant solutions are also clarified, as we find two replicae in the chiral limit, with one of them collapsing if a small quark mass is added, leaving only one replica. This is of course contingent on the strength of the interaction model, and we have performed computations with the standard Cornell strength of the interquark potential when applied to a well-known field theory with the global quantum numbers of QCD (Quantum Chromodynamics.
(In a previous harmonic oscillator analysis~\cite{Bicudo:2019ryc} some of us  found a tower of replicae, possibly infinite, but with energies not necessarily below the perturbative vacuum state.)

We work in the equal-time quantization formalism. It is known that, although widely used in sum rule investigations~\cite{Gubler:2018ctz}, in other formalisms such as light--front quantization, 
a quark condensate is not a natural concept,  and it has been proposed that its appearance is restricted to hadrons~\cite{Brodsky:2012ku}
and not to the full space. Our approach in this is traditional~\cite{Nambu:1961tp}, 
the condensate being invariant under translations. 

The BCS formalism applied to a  Hamiltonian theory, in the spirit of the NJL (Nambu-Jona-Lasinio) model but for generic nonlocal two-body interactions is quickly reviewed to settle the necessary notation (see Sec.~\ref{sec:BCS}).

In this formalism, different vacua are exactly orthogonal in an infinite volume, and have exponentially suppressed overlaps for volumes above $\sim (4\pi/3)(2.5 {\rm fm})^3$.
This gives rise to theory conundrums in which we do not wish to delve, but a quick summary, for context, is presented in section~\ref{sec:inequivalent}. 
Suffice it to say that a complete transition over all space volume from one to another vacuum would require cosmological times.

This BCS theory is then applied in Sec.~\ref{sec:model} to the specific Hamiltonian of the North Carolina State University group, a field theory featuring a Cornell (funnel or Coulomb+linear) potential. 

Our numerical solutions to the BCS gap equation (the well-known ground-state one dating to Adler and Davies~\cite{Adler:1984ri} as well as the replicae) are presented in Sec.~\ref{sec:solutions}.

Further new content is presented in Sec.~\ref{sec:level4}
 where we discuss the replicae as saddle points of the energy density, and find the number of negative eigenvalues over each of the relevant vacua; from this analysis we can confidently conjecture that we have already found all the possible 
vacua of this model (and, since the interaction strength is typical of the strong interactions, it is dubious that QCD would have many more than the ones at hand). 

We can numerically slice the energy density surfaces in the (infinite-dimensional) BCS gap-angle function space to visually show (Sec.~\ref{sec:slices}) the relative placement of the different states discussed, confirming their ordering.

In Sec.~\ref{sec:mesons} we show how the spectrum of hadrons over the excited vacua is not that different from the spectrum over the ground-state vacuum, which is inconvenient to discern on which state the actual hadrons are built. 

Whereas the spectrum, at the model level of precision, is not reliable to distinguish the replica, the hadron wavefunctions (and condensate) are much more sensitive functions of the gap angle and therefore of the choice of vacuum state, as shown in Sec.~\ref{sec:condensateandfpi}, and can be used to distinguish the replicae and ascertain that traditional hadron physics is closest in order of magnitude to the ground-state BCS vacuum, even with all its attending deficiencies.

Seeing that we have exhausted our model-based insights, we propose, in section~\ref{sec:latticesearch}, a strategy for further work to see whether full QCD as treated on the lattice can produce the replicae given the difficulty to study them in experiment without further model-independent information.

Sec.~\ref{sec:conclusions} then wraps the discussion up.

A short note based on this work has been presented at an international conference~\cite{Bicudo:2021vrm}. The present article constitutes the 
full write up of the investigation.

%%%%%%%%%%%%%%%%%%%%%%%%%%%%%%%%%%%%%%%%%%%%%%%%%%%%%%%%%%%%%%%%%%%%%%%%%%%%%%%%%%%%%%%%%%%%%%%%%%%%
\section{Chiral vacuum in BCS approximation and possibility of excited states} \label{sec:BCS}
%%%%%%%%%%%%%%%%%%%%%%%%%%%%%%%%%%%%%%%%%%%%%%%%%%%%%%%%%%%%%%%%%%%%%%%%%%%%%%%%%%%%%%%%%%%%%%%%%%%%

The ground state of the interaction--free quark Hamiltonian $\arrowvert 0 \rangle$ is not the ground state of a fully interacting Hamiltonian, that we can denote by $|\Omega\rangle$.
 
In terms of the degrees of freedom in which the Hamiltonian is formulated, the theory has a non-vanishing quark condensate $\langle\Omega|\Bar{\Psi} \Psi|\Omega\rangle \neq 0$, formed by quark-antiquark $^3P_0$ Cooper pairs (in analogy to superconductor theory). 
So, following with the analogy, it is standard to use the BCS many-body technique to variationally approximate the non-chiral ground state (or simply BCS vacuum) $|\Omega\rangle$. 

A slick way of writing it down starts from the momentum expansion of the quark field (for another basis, see~\cite{Yepez-Martinez:2021wmb,Yepez-Martinez:2018anj})
\begin{equation}
    \Psi(\vec{x})=\sum_{c\lambda}\!\int\!\dfrac{\dif^3 k}{(2\pi)^3}\!\left[u_{c\lambda}(\Vec{k})b_{c\lambda}(\Vec{k})+v_{c\lambda}(-\Vec{k})d^\dagger_{c\lambda}(-\Vec{k})\right]\! e^{i\Vec{k}\cdot \Vec{x}}
\end{equation}
with $u_{c\lambda}$, $v_{c\lambda}$ the bare particle and antiparticle Dirac spinors, $b_{c\lambda}$, $d_{c\lambda}$ the bare particle, antiparticle annihilation operators, $\lambda$ the spin state and $c=1,2,3$ the color index (mostly suppressed in what follows). 

We can expand $\Psi$ in terms of any complete basis, so we choose to expand it using a new quasiparticle basis
\begin{equation}
    \Psi(\vec{x})=\sum_{\lambda}\!\int\!\dfrac{\dif^3 k}{(2\pi)^3}\!\left[U_{\lambda}(\Vec{k})B_{\lambda}(\Vec{k})+V_{\lambda}(-\Vec{k})D^\dagger_{\lambda}(-\Vec{k})\right]\! e^{i\Vec{k}\cdot \Vec{x}}.
\end{equation}
The two bases are related by a linear  Bogoliubov-Valatin transformation (see \cite{Fetter}, \cite{rin80} for an in-depth treatment),
\begin{align} \label{Bogoliubov1}
    &B_\lambda(\Vec{k})= \alpha_k b_\lambda (\Vec{k})-\beta_kd^\dagger_\lambda (\Vec{k}),\\
    &D_\lambda(-\Vec{k})=\alpha_kd_\lambda (-\Vec{k})+\beta_k b^\dagger_\lambda (\Vec{k}).
\end{align}
The coefficients $\alpha_k$, $\beta_k$ only depend on $|\Vec{k}|$,  and are real $c$-numbers. This linear transformation is canonical if and only if the new operators obey the same anticommutation relations than the original ones 
\begin{align}
    \{B_k,B^\dagger_{k'}\}=\{D_k,D^\dagger_{k'}\}=\delta_{kk'}.
\end{align}
This implies $|\alpha_k|^2+|\beta_k|^2=1$ and we can implement this transformation as a rotation, parametrized by a Bogoliubov angle $\theta(k)\equiv\theta_k$, a function of $k$. This parametrization yields the following relation between the two bases from Eq.~(\ref{Bogoliubov1}),
\begin{align}
    \label{Bogoliubov} &B_\lambda(\Vec{k})=\cos \frac{\theta_k}{2}b_\lambda (\Vec{k})-\lambda\sin \frac{\theta_k}{2}d^\dagger_\lambda (\Vec{k}), \nonumber  \\
    &D_\lambda(-\Vec{k})=\cos \frac{\theta_k}{2}d_\lambda (-\Vec{k})+\lambda \sin \frac{\theta_k}{2}b^\dagger_\lambda (\Vec{k}).
\end{align}
It is often more convenient to work in terms of the BCS gap angle $\phi_k\equiv\phi(k)$, which is related to the earlier one by $\phi=\theta+\alpha$, where $\alpha$ is the perturbative mass angle satisfying, (in terms of the quark mass $m_q$ and  $E_k=\sqrt{m_q^2+k^2}$, the energy at momentum $k$)  $\sin \alpha=m_q/E_k$. Then the rotated quasiparticle spinors can be expressed in terms of the original spinors as follows
\begin{align}
\label{eq:U}
\begin{split}
    U_\lambda(\Vec{k})&=\cos \dfrac{\theta_k}{2} u_\lambda(\Vec{k})-\lambda\sin \dfrac{\theta}{2}v_\lambda(-\Vec{k})\\
    &=\dfrac{1}{\sqrt{2}}
        \renewcommand*{\arraystretch}{1.5}
    \begin{pmatrix}
    \sqrt{1+\sin \phi_k} \, \xi_\lambda\\
     \sqrt{1-\sin \phi_k}\, \Vec{\sigma}\cdot \hat{k} \, \xi_\lambda
    \end{pmatrix},
\end{split}\\
\begin{split}
    V_\lambda(-\Vec{k})&=\cos \dfrac{\theta_k}{2} v_\lambda(-\Vec{k})+\lambda\sin \dfrac{\theta}{2}u_\lambda(\Vec{k})\\
    &=\dfrac{1}{\sqrt{2}}
        \renewcommand*{\arraystretch}{1.5}
    \begin{pmatrix}
         -\sqrt{1-\sin \phi_k}\, \Vec{\sigma}\cdot \hat{k} \,i\sigma_2 \,\xi_\lambda\\
    \sqrt{1+\sin \phi_k} \,i\sigma_2\, \xi_\lambda
    \end{pmatrix}
\end{split}
\end{align}
with $\xi_\lambda$ a two-dimensional Pauli spinor. 

The trivial perturbative vacuum, defined by $b_\lambda|0\rangle=d_\lambda|0\rangle=0$, is related to the BCS vacuum, defined by $B_\lambda|\Omega\rangle=D_\lambda|\Omega\rangle=0$, by the transformation
\begin{equation}
\label{eq:vacuum}
\begin{split}
    |\Omega\rangle=\exp \Bigg(-\sum_{\lambda_1,\lambda_2} \int \dfrac{\dif^3 k}{(2\pi)^3} (\vec{\sigma}\cdot \hat{k})_{\lambda_1 \lambda_2} \tan{\dfrac{\theta_k}{2}}\\
    \times b_{\lambda_1}^\dagger(\Vec{k})
    d_{\lambda_2}^\dagger(-\Vec{k}) \Bigg)|0\rangle
    \end{split}
\end{equation}
up to an important normalization that will come to the fore in the next Sec.~\ref{sec:inequivalent}.

If we expand the exponential, we see the operators $b^\dagger d^\dagger$ create a current quark-antiquark pair, exposing the BCS vacuum as a coherent state of $q\Bar{q}$ excitations producing a $^3{P}_0$ condensate [notice the $\vec{\sigma}\hat{\cdot} k$ factor in Eq.~(\ref{eq:vacuum})]. 

To find the optimal $\arrowvert \Omega \rangle$ vacuum, we minimize the expectation value of the energy density, 
using the gap angle $\phi_k$ as a variational function,
\begin{eqnarray}
\label{1der}
     \dfrac{\delta \langle \Omega| H |\Omega\rangle}{\delta \phi_k}=0.
\end{eqnarray}
This leads to a dynamical mass gap equation that is nonlinear and will need to be numerically solved. 
A model implementation will be shown shortly, in Eq.~(\ref{eq:massgap}).
The solutions $\phi^i(k)\equiv\phi_k^i$ of this equation, substituted with $\theta_k=\phi_k-\arctan(m_q/k)$ in Eq.~(\ref{eq:vacuum}), are interpreted as possible vacua of the theory, so the numerical task is to solve this nonlinear integral equation. Its nonlinear character is precisely what allows the existence of other solutions besides the BCS vacuum.

If the interactions are turned off and only the free Hamiltonian (mass and kinetic terms for the quarks) is considered,
the gap equation trivially returns $\phi = 0$ in the massless case, which continuously deforms into a vacuum gap function
with $\sin\phi (k) = m_q/\sqrt{m_q^2+k^2}$ for finite mass. This we will call the ``perturbative'' vacuum (in $S$-matrix theory it is often named the ``asymptotic'' vacuum instead). 

The number and nature of the solutions of the gap equation is unknown {\it a priori} and needs to be found through 
numerical exploration.

%%%%%%%%%%%%%%%%%%%%%%%%%%%%%%%%%%%%%%%%%%%%%%%%%%%%%%%%%%%%%%%%%%%%%%%%%%%%%%%%%%%%%%%%%%%%%%%%%%%%
\section{Inequivalent representations  of the operator algebra }\label{sec:inequivalent}
%%%%%%%%%%%%%%%%%%%%%%%%%%%%%%%%%%%%%%%%%%%%%%%%%%%%%%%%%%%%%%%%%%%%%%%%%%%%%%%%%%%%%%%%%%%%%%%%%%%%

%%%%%%%%%%%%%%%%%%%%%%%%%%%%%%%%%%%%%%%%%%%%%%%%%%%%%%%%%%%%%%%%%%%%%%%%%%%%%%%%%%%%%%%%%%%%%%%%%%%%
\subsection{Orthogonality in the infinite-volume limit}
%%%%%%%%%%%%%%%%%%%%%%%%%%%%%%%%%%%%%%%%%%%%%%%%%%%%%%%%%%%%%%%%%%%%%%%%%%%%%%%%%%%%%%%%%%%%%%%%%%%%
In this section we take notice that, in the infinite volume limit, the trivial vacuum and the BCS vacuum are orthogonal $\langle \Omega|0\rangle=0$. This observation extends also to the replicae.
Moreover, the entire Fock spaces built over each of the vacua are mutually orthogonal.
This is typical of infinitely-many body problems as discussed in early literature of the field~\cite{Barton:1963zi}.

In a finite volume there exists a unitary operator $U$ that effects the transformation of Eq.~(\ref{Bogoliubov}) on the states, $|n'\rangle = U | n\rangle$.
The states of the type $|n\rangle$ constructed by applying $b^\dagger$, $d^\dagger$ over $|0\rangle$
and those of type $|n'\rangle$ from applying $B^\dagger$, $D^\dagger$ over $|\Omega \rangle$ lead to typical overlaps of the form
(from Eq.~(\ref{Bogoliubov}) and from Eq.~(13.50) in~\cite{Barton:1963zi})
\begin{eqnarray} \label{nulloverlap}
\langle n' | n \rangle \propto 
\langle \Omega | 0 \rangle &\propto& \prod_{k,\lambda} \cos\frac{\theta_k}{2}
\nonumber \\ &\propto& 
\exp\left(
2\sum_k \log (\cos\frac{\theta_k}{2})
\right)
\nonumber \\
&\propto & \exp\left( V\frac{1}{\pi^2}
\int k^2 \log(\cos\frac{\theta_k}{2}) dk
\right) \nonumber \\
& \propto & \exp(-V\times |c|)
\end{eqnarray}
in view of the continuum-limit correspondence 
$\sum_k \to \frac{V}{((2\pi)^3}\int d^3k$.

That overlap in Eq.~(\ref{nulloverlap}) falls exponentially with volume as shown in the last line, and quickly vanishes, for every $\theta_k/2$ Bogoliubov profile barring $\theta_k=0$ which makes the constant $|c|$ multiplying $V$ in the exponent become null: this is just the noninteracting case $|\Omega\rangle = |0\rangle$ where there is no Bogoliubov rotation and all vacua are the same as the perturbative trivial one.

Another way to understand this orthogonality at the operator level is by noting that the relation $B_k = U(\theta) b_k U^{-1}(\theta)$ which would effect the Bogoliubov transformation from the bare to the dressed quark operators requires~\cite{Blasone:2017spx}
\begin{eqnarray}
    U(\theta) &=& \exp  \left(-\sum_\lambda\int \frac{d^3k}{(2\pi^3)} \frac{\theta_k}{2}
    (b_{k\lambda}d_{-k\lambda}-b^\dagger_{k\lambda} d^\dagger_{-k\lambda})\right) \nonumber \\
    &=& \exp\left(-2V \int \frac{d^3k}{(2\pi^3)} \log\cos\frac{\theta_k}{2} \right) \times \nonumber \\ & &  \exp\left(-\sum_\lambda\int \frac{d^3k}{(2\pi^3)}\tan \frac{\theta_k}{2} b^\dagger_{k\lambda} d^\dagger_{-k\lambda} \right)  
    \times \nonumber \\ & &
    \exp \left( -\sum_\lambda\int \frac{d^3k}{(2\pi^3)}\tan \frac{\theta_k}{2} b_{\lambda}k d_{-k\lambda} \right)  
\end{eqnarray}
with the first exponential after the last equality carrying $V\to \delta^{(3)}(0)$ which drives the transformed state
$|\Omega\rangle = U(\theta) |0\rangle$ to have zero overlap with the initial one upon taken the infinite volume limit (the operator $U(\theta)$ is improperly unitary).

\begin{figure} \centering
\includegraphics[width=0.8\columnwidth]{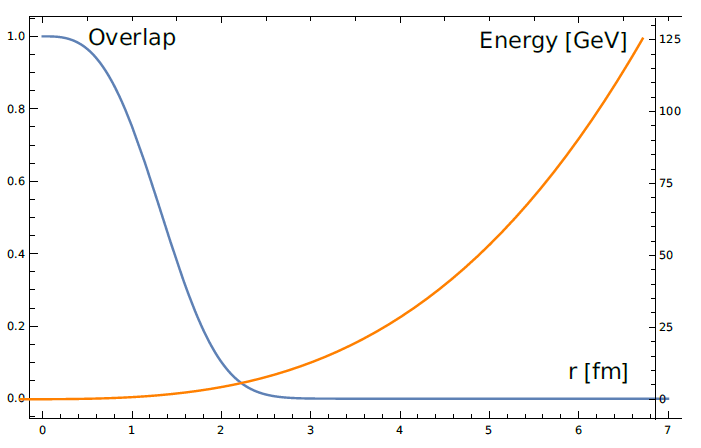}
\caption{\label{fig:overlap} The first vacuum replica 
has an energy
over the BCS vacuum, the zero point, which grows linearly with the volume, hence with the cube of the radius in a spherical bubble. Simultaneously, its overlap with the ground state $\langle \Omega | \Omega' \rangle$
drops exponentially with the quantization volume, and for radii larger than 2.5 fermi it becomes negligible.  }
\end{figure}

The situation is depicted in figure~\ref{fig:overlap}.
It displays the energy of a bubble of replica vacuum (above the ground state BCS vacuum) as function of the radius of the said bubble, together with the overlap of both vacua states quantized over the volume of the bubble alone, so that $\exp(-V)$ remains finite.
It is clear from the figure that for radii above $2.5$ fermi the overlap is significantly smaller. The precise number depends on the details of the model used, and we will soon turn to modeling in Sec.~\ref{sec:model}.

As discussed by Barton~\cite{Barton:1963zi}, there is a symmetry reason why, if we accept say $|0\rangle$ or alternatively $|\Omega\rangle$ as having norm 1, the other one, and also any other potential vacua such as the $|\Omega'\rangle$ replica cannot have the same norm. 

Instead, the false vacuum states built over the ground state one belong to the continuous spectrum of the momentum operator and are subject to delta-distribution normalization $\langle {\bf p} | {\bf p'} \rangle = \delta^{(3)}({\bf p}-{\bf p'})$; since translation invariance for an adequate vacuum state requires ${\bf p}={\bf 0}$, we see a volume ($\delta^{(3)}(0)$) divergence in the normalization. No translation invariance states normalizable to 1 are available for mixing with the initially chosen ground state. The excited vacua generate disjoint or inequivalent copies of the entire Fock space.

%%%%%%%%%%%%%%%%%%%%%%%%%%%%%%%%%%%%%%%%%%%%%%%%%%%%%%%%%%%%%%%%%%%%%%%%%%%%%%%%%%%%%%%%%%%%%%%%%%%%
\subsection{Semiclassical treatment of transitions}
%%%%%%%%%%%%%%%%%%%%%%%%%%%%%%%%%%%%%%%%%%%%%%%%%%%%%%%%%%%%%%%%%%%%%%%%%%%%%%%%%%%%%%%%%%%%%%%%%%%%

No fully nonperturbative and quantum solution to the issue of orthogonality of the Fock spaces generated by improperly unitary transformations such as the Bogoliubov rotation is known to us. This means that, in practice, total transitions $|\Omega\rangle\to |\Omega'\rangle$ among any of the states here discussed require a time exponentially diverging with the volume.

Phenomenologically, that orthogonality raises problems with transitions that actually occur in the lab and that are believed to have taken place in the early universe.
For example, solids can transition from the superconducting state (that is in the BCS-paired state) to 
the normal state with finite resistivity. But if no operator connects the two states, 
how is the transition to be phenomenologically described?

An accepted working recipe to address transitions between orthogonal vacua in systems with infinitely many degrees of freedom was proposed by Coleman~\cite{Coleman:1977py}. 
In a nutshell, one closes the eyes to the fact that no operator connects the two inequivalent Fock spaces and proceeds in a finite volume where the transition is permitted. Then, the walls of this finite--volume bubble expand with speed $c$ transforming more of the false vacuum into the true one (and simultaneously, any of the excited particles over the false vacuum into particles over the true vacuum). Semiclassical equations can be written for this false to true vacuum transition.

In what concerns us here, this way of thinking means that a false vacuum characterized by a higher energy density than the ground state can classically roll towards it (if such false vacuum is a saddle point with directions in $\theta_k$ function space that have negative second derivative of the energy density) or tunnel to it (if the false vacuum is metastable) over finite volumes. 
We will numerically find below that, at least in the model of QCD which we are handling here, the first case is realized and the false vacuum is unstable in the classical sense, to collective Bogoliubov rotations; but metastable to few-body excitations, with hadrons built over the vacuum having positive mass.
Considering the entire spatial volume, the transition is never really completed: one evolves over a continuum of interpolating states which have one of the vacuum-wavefunction angles $\phi$ over a certain volume and another $\phi'$ over the rest.

%%%%%%%%%%%%%%%%%%%%%%%%%%%%%%%%%%%%%%%%%%%%%%%%%%%%%%%%%%%%%%%%%%%%%%%%%%%%%%%%%%%%%%%%%%%%%%%%%%%%
\section{Model implementation: global--color Hamiltonian field theory with a Cornell-like potential}
\label{sec:model}
%%%%%%%%%%%%%%%%%%%%%%%%%%%%%%%%%%%%%%%%%%%%%%%%%%%%%%%%%%%%%%%%%%%%%%%%%%%%%%%%%%%%%%%%%%%%%%%%%%%%
The discussion is illustrated by  QCD in the Coulomb gauge, that can be modeled along the lines of \cite{Llanes-Estrada:2004edu},
that features a Cornell-like model for the longitudinal color interaction and a Yukawa-type gluon exchange for the transverse-gluon mediated interactions.
This is based on earlier work by several groups~\cite{RodriguezMarrero:1984zt,LeYaouanc:1984ntu,Szczepaniak:1996gb,Robertson:1998va} etc.
The QCD Coulomb gauge Hamiltonian
\begin{eqnarray}
    H=H_q+H_g+H_{qg}+V_C,
\end{eqnarray}
with
\begin{align}
    &H_q=\int \dif^3 x\,  \Psi (\Vec{x})^\dagger(-i\vec{\alpha}\cdot \vec{\nabla}+\beta m_q)\Psi(\Vec{x}),\\
    &H_g=\text{Tr}\int \dif^3 x\,  [\Vec{\Pi}^a(\vec{x})\cdot \Vec{\Pi}^a(\vec{x})+\Vec{B}^a(\vec{x})\cdot \Vec{B}^a(\vec{x})],\\
    &H_{qg}=g \int \dif^3 x\, \Vec{J}^a(\Vec{x})\cdot \Vec{A}^a(\vec{x}),\\
    \label{eq:potc}
    &V_C=-\dfrac{1}{2}\int \dif^3 x \dif^3 y\,  \rho^a(\vec{x})V(|\vec{x}-\vec{y}\,|) \rho^a(\vec{y}),
\end{align}
is simplified into a model Hamiltonian by reducing the field-theoretical kernel to a $c$-function potential $V$. 
Here, $\Psi$ and $m_q$ are the (bare) quark field and mass, $\rho^a(\vec{x})=\Psi^\dagger_x T^a \Psi_x$ and $\Vec{J}^a(\vec{x})=\Psi^\dagger_x T^a\Vec{\alpha} \Psi_x$ are  respectively the colour density and current, with $T^a$ the generators of SU(3), $g$ is the QCD coupling, $\Vec{A}^a$ are the gauge fields, $\Vec{\Pi}^a$ are the conjugate fields and $\Vec{B}^a$ are the non-abelian magnetic fields defined by
\begin{eqnarray}
  \Vec{B}^a \equiv \Vec{\nabla} \times \Vec{A}^a+\dfrac{1}{2}f^{abc} \Vec{A}^b \times \Vec{A}^c
\end{eqnarray}
with $f^{abc}$ the structure constants of SU(3). 
For a complete analysis of the Coulomb gauge Hamiltonian see \cite{Christ:1980ku}. 

Let us focus on the potential $V_C$ of Eq.~(\ref{eq:potc}). A model of the strong interaction should reflect the phenomenon of confinement, that is, the absence of isolated color charged particles (such as gluons or quarks) in the spectrum. In this work, we model confinement through the potential $V_C$. The $V_C$ part is taken as a Cornell potential, i.e. a Coulomb potential due to the exchange of a gluon plus a linear part which is responsible for confinement:
\begin{equation}
    V_{\text{Cornell}}(k)=-4\pi \dfrac{\alpha_s}{k^2}-8\pi\dfrac{\sigma}{k^4}
\end{equation}
where $\alpha_s$ is the coupling in QCD and $\sigma$ is a string tension constant, which can be inferred from experimental data (for example, of charmonium spectrum) in lattice QCD calculations.

In particular, we use a modified Cornell potential numerically fitted to pure Yang-Mills variational computations (see \cite{Szczepaniak:2001rg})
\begin{equation} \label{Cornellpotential}
    V_C(p)=\begin{cases}
    C(p)\equiv -\dfrac{8.07}{p^2}\dfrac{\log^{-0.62}\left(\frac{p^2}{m_g^2}+0.82\right)}{\log^{0.8}\left(\frac{p^2}{m_g^2}+1.41\right)} \hspace{0.2cm} &\text{if} \hspace{0.1cm} p>m_g\\
    L(p)\equiv -\dfrac{12.25\, m^{1.93}_g}{p^{3.93}} \hspace{0.2cm} &\text{if} \hspace{0.1cm} p<m_g
    \end{cases}
\end{equation}
where the low momentum component is (numerically) close to a pure linear potential and the other term represents a renormalized high energy Coulomb tail.

\begin{figure}[h]
\centering
\includegraphics[width=0.7\columnwidth]{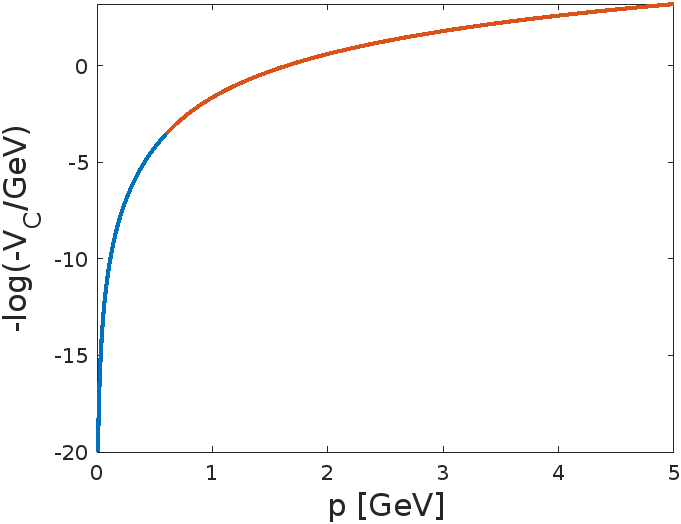}
\includegraphics[width=0.7\columnwidth]{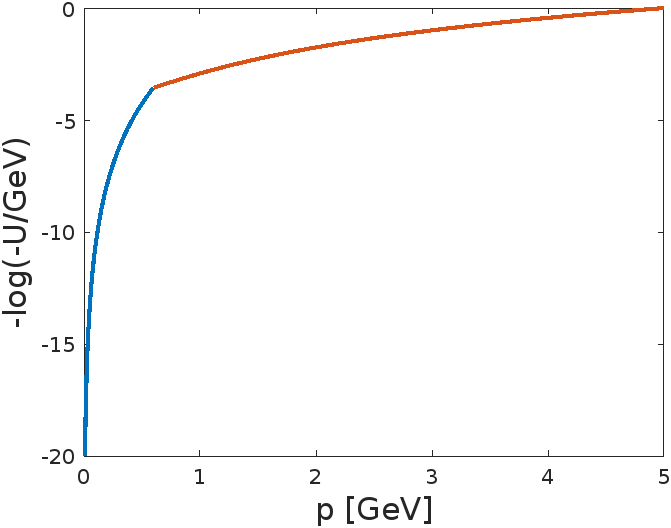}
\caption{\label{fig:potentials} The upper plot represents the Coulomb-like potential (Cornell potential from Eq.~(\ref{Cornellpotential})) in momentum space, 
while the lower plot is the transverse one from Eq.~(\ref{eq:hyper}), matching a Coulombic and a Yukawa like tail that represents massive gluon exchange.}
\end{figure}

Because chiral symmetry is a feature of quarks, not gluons, $H_g$ is omitted and we substitute $H_{qg}$ by a generic transverse hyperfine interaction $V_T$ due to the exchange of a transverse gluon, according to
\begin{align}
\label{eq:hyper}
      &V_T=\dfrac{1}{2}\int \dif^3 x \dif^3 y\, \Vec{J}^a_i(\vec{x})U_{ij}(\vec{x},\vec{y}) \Vec{J}^a_j(\vec{y}),\\ 
    &U_{ij}(\vec{x},\vec{y})=\left(\delta_{ij}-\dfrac{\nabla_i \nabla_j}{\nabla^2}\right)_{\Vec{x}}U(|\vec{x}-\vec{y}\,|),\\
    &U(p)=\begin{cases}
    C(p) \hspace{0.1cm} &\text{if} \hspace{0.2cm} p>m_g\\
    -\dfrac{C_h}{p^2+m^2_g} \hspace{0.2cm} &\text{if} \hspace{0.1cm} p<m_g
    \end{cases}
\end{align}
with a Yukawa type interaction at low momentum and $C_h$ a constant determined by matching the high and low momentum regions. This interaction is sensible for transferred momenta not much larger than the dynamical mass of the gluon $m_g\sim 600$ MeV, which we are using, in practice, as the scale of the theory. 
Both longitudinal and transverse gluon potentials in momentum space are plot in figure~\ref{fig:potentials}.

To find the possible vacua of this theory, the extremal points of the Hamiltonian  expected value 
from Eq.~(\ref{1der}), we employ the model Hamiltonian yielding
%\begin{widetext}
\begin{equation}
\begin{alignedat}{3}
\label{eq:rho}
    \rho\equiv&\dfrac{\langle \Omega| H |\Omega\rangle}{V}= \int \dfrac{\dif^3 k}{(2\pi)^3}\big[-6(k c_k+m_q s_k)\\
    &-2\int \dfrac{\dif^3 q}{(2\pi)^3} \hat{V}(|\Vec{k}-\Vec{q}\,|)(1-s_k s_q -c_k c_q x)\\
    &+4\int\dfrac{\dif^3 q}{(2\pi)^3}\hat{U}(|\Vec{k}-\Vec{q}\,|)(1+s_k s_q)+c_q c_k\hat{W}(|\Vec{k}-\Vec{q}\,|)\big]
\end{alignedat}
\end{equation}
%\end{widetext}
where $\rho$ is the energy density of the system such that $\int \dif^3 x\, \rho=H$, $s_k\equiv \sin \phi(\Vec{k})$, $c_k\equiv \cos \phi(\Vec{k})$ and
\begin{equation}
\label{eq:W}
  \hat{W}(|\Vec{k}-\Vec{q}\,|) \equiv\dfrac{x(|\Vec{k}|^2+|\Vec{q}\,|^2)-|\Vec{k}||\Vec{q}\,|(1+x^2)}{|\Vec{k}-\Vec{q}\,|^2} \hat{U}(|\Vec{k}-\Vec{q}\,|)
\end{equation}
with $x=\hat{k}\cdot\hat{q}$.

The mass gap equation is then
%\begin{widetext}
\begin{equation}
\begin{split}
\label{eq:massgap}
    k s_k-&m_qc_k=\dfrac{2}{3}\int \dfrac{\dif^3 q}{(2\pi)^3} \hat{V}(|\Vec{k}-\Vec{q}\,|)[s_k c_q x-c_k s_q]\\
    &-\dfrac{4}{3}\int\dfrac{\dif^3 q}{(2\pi)^3}\big(c_k s_q\hat{U}(|\Vec{k}-\Vec{q}\,|)-c_q s_k \hat{W}(|\Vec{k}-\Vec{q}\,|)\big).
\end{split}
\end{equation}
%\end{widetext}

The angular integrals of the kernel can be separately  evaluated 
\begin{equation}
\begin{split}
    k s_k-m_qc_k&=\dfrac{1}{6\pi^2}\int_0^\infty \dif q\,q^2[ s_k c_q V_1-c_k s_q V_0\\
    &-2(c_k s_qU_0 -c_q s_k W_0)]
\end{split}
\end{equation}
where
\begin{eqnarray}
\label{eq:angulars}
  I_n=\int_{-1}^1 \dif x\, I(|\vec{k}-\vec{q}|) x^n
\end{eqnarray}
with $I=V,U,W$ from Eqs.~(\ref{eq:rho}) and (\ref{eq:W}). Let us first consider the chiral limit $m_q=0$. In this limit, it is straightforward to check that we have the trivial solution $\sin \phi_k=0$. This solution is clearly the perturbative vacuum $|0\rangle$ (there is no Bogoliubov rotation). The non-trivial solutions on the other hand need to be found numerically. In addition, in this limit $m_q \rightarrow 0$ the mass gap equation is symmetric under the exchange of the sign of ($\sin \phi_k$), that is, if $\sin \phi_k$ is a solution, then $-\sin \phi_k$ is a solution as well. We break this ambiguity by choosing, as need arises, $\sin \phi_k \to 0^+$ to be positive at either very large or very small momentum.

%%%%%%%%%%%%%%%%%%%%%%%%%%%%%%%%%%%%%%%%%%%%%%%%%%%%%%%%%%%%%%%
\section{Numerical solutions: BCS angles and quark masses \label{sec:solutions}}
%%%%%%%%%%%%%%%%%%%%%%%%%%%%%%%%%%%%%%%%%%%%%%%%%%%%%%%%%%%%%%%

Three solutions $\phi^0_k$, $\phi^1_k$, $\phi^2_k$ to the discretized gap equation in the chiral limit $m_q\to 0$ are numerically found  by linear iteration of the system with Newton's method, as described elsewhere~\cite{Llanes-Estrada:2001bgw}

\begin{figure}[h]
\includegraphics[scale=0.2]{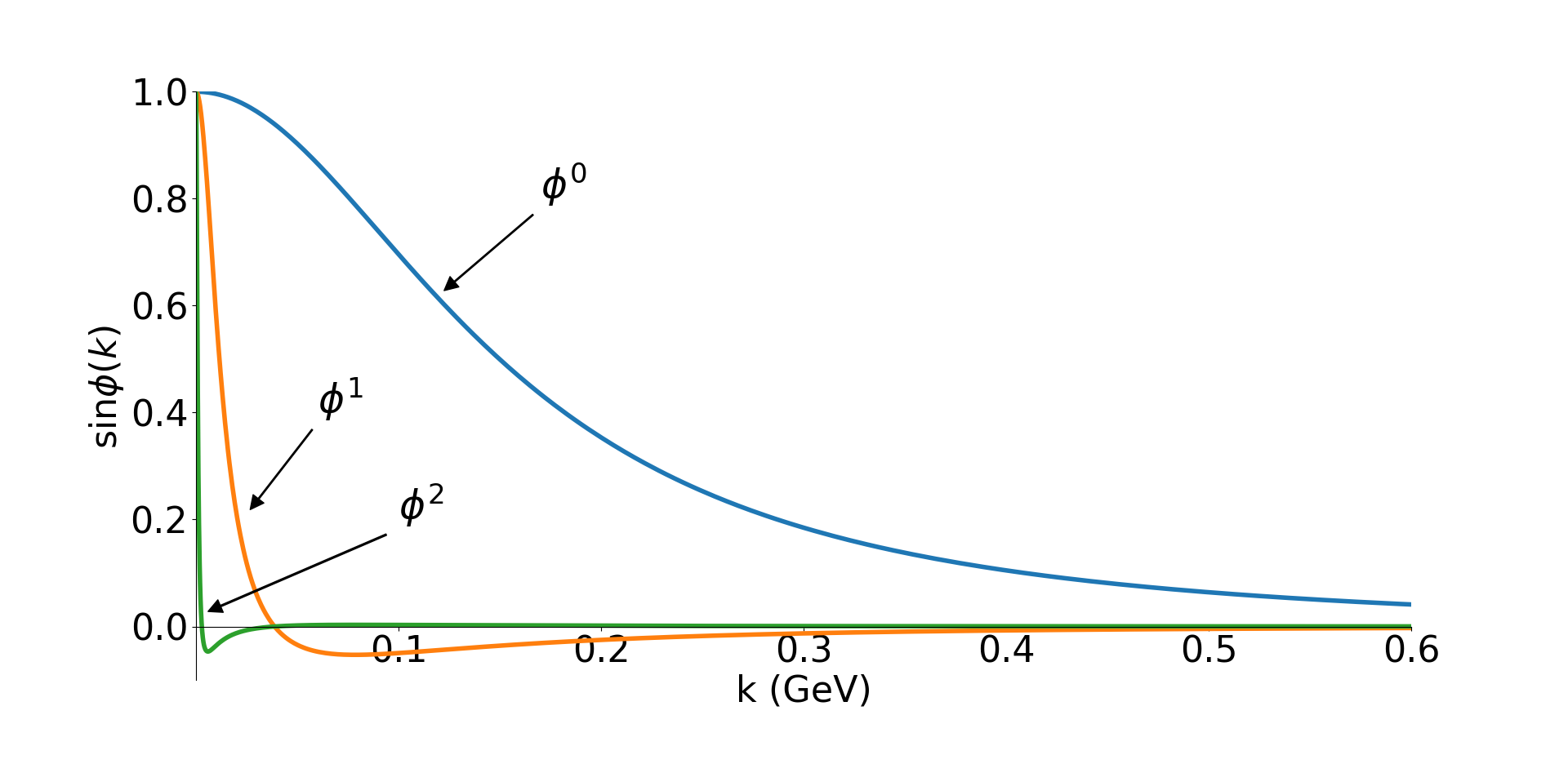}
\caption{\label{fig:sols0} The first three (non-trivial) solutions of the mass gap equation in the chiral limit $m_q=0$. We have analogous solutions under the exchange $\sin \phi_k  \leftrightarrow -\sin\phi_k$, but we choose here only those in the first quadrant for small $k$ (the other ones, in the massive case, appear only for unphysical $m_q<0$).}
\end{figure}

\begin{figure}[h]
\includegraphics[scale=0.18]{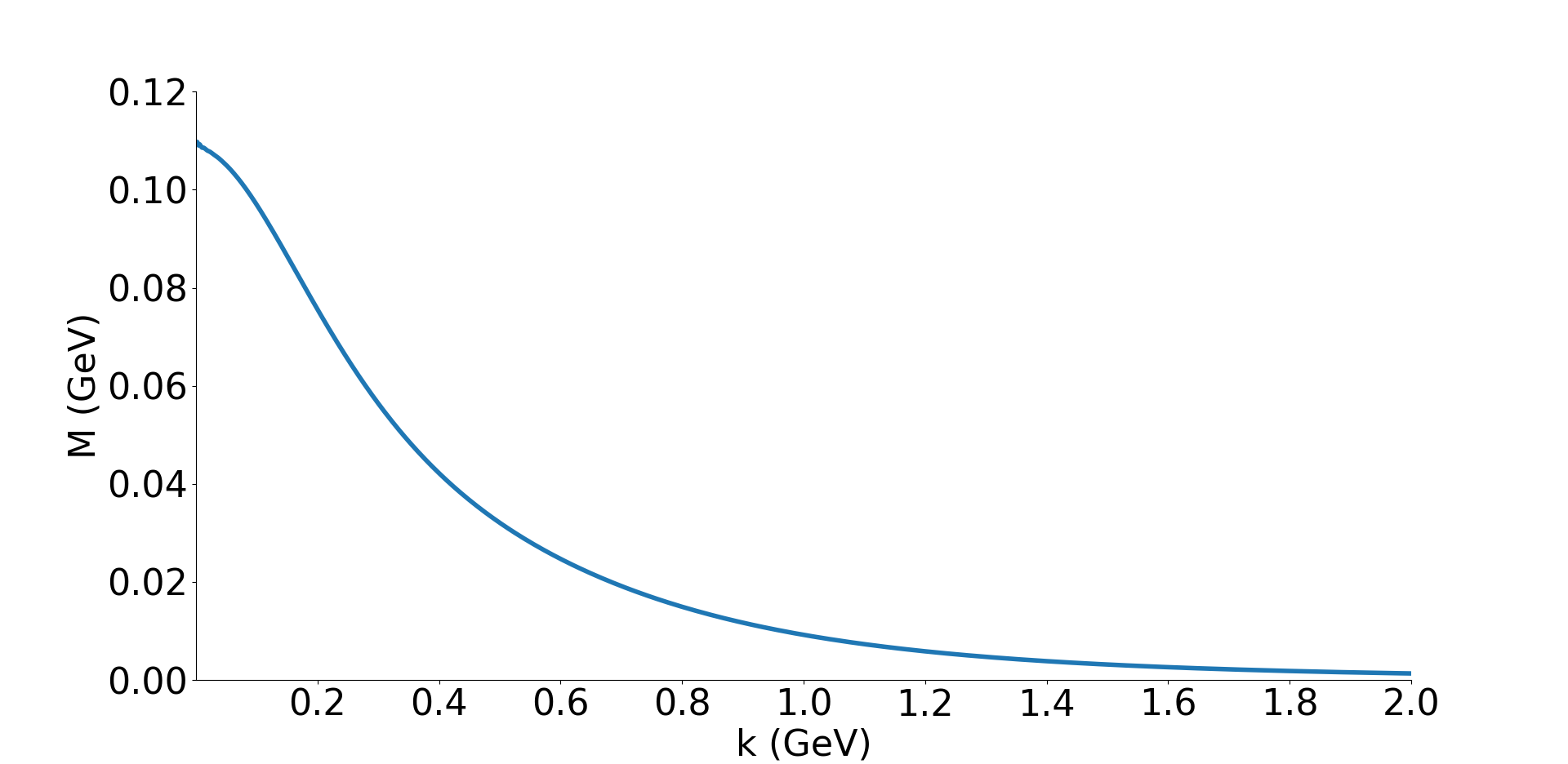}
\includegraphics[scale=0.18]{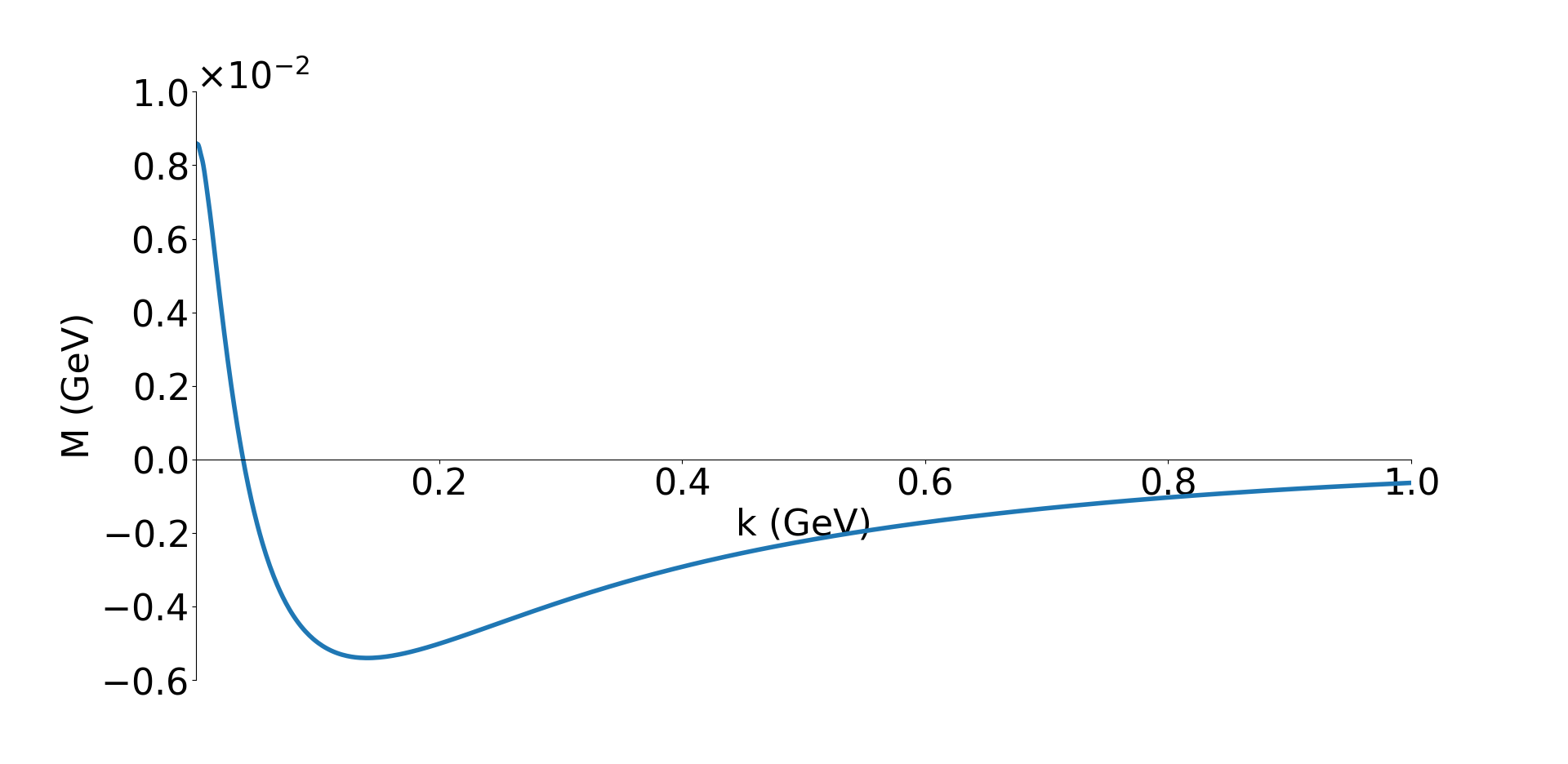} 
\includegraphics[scale=0.18]{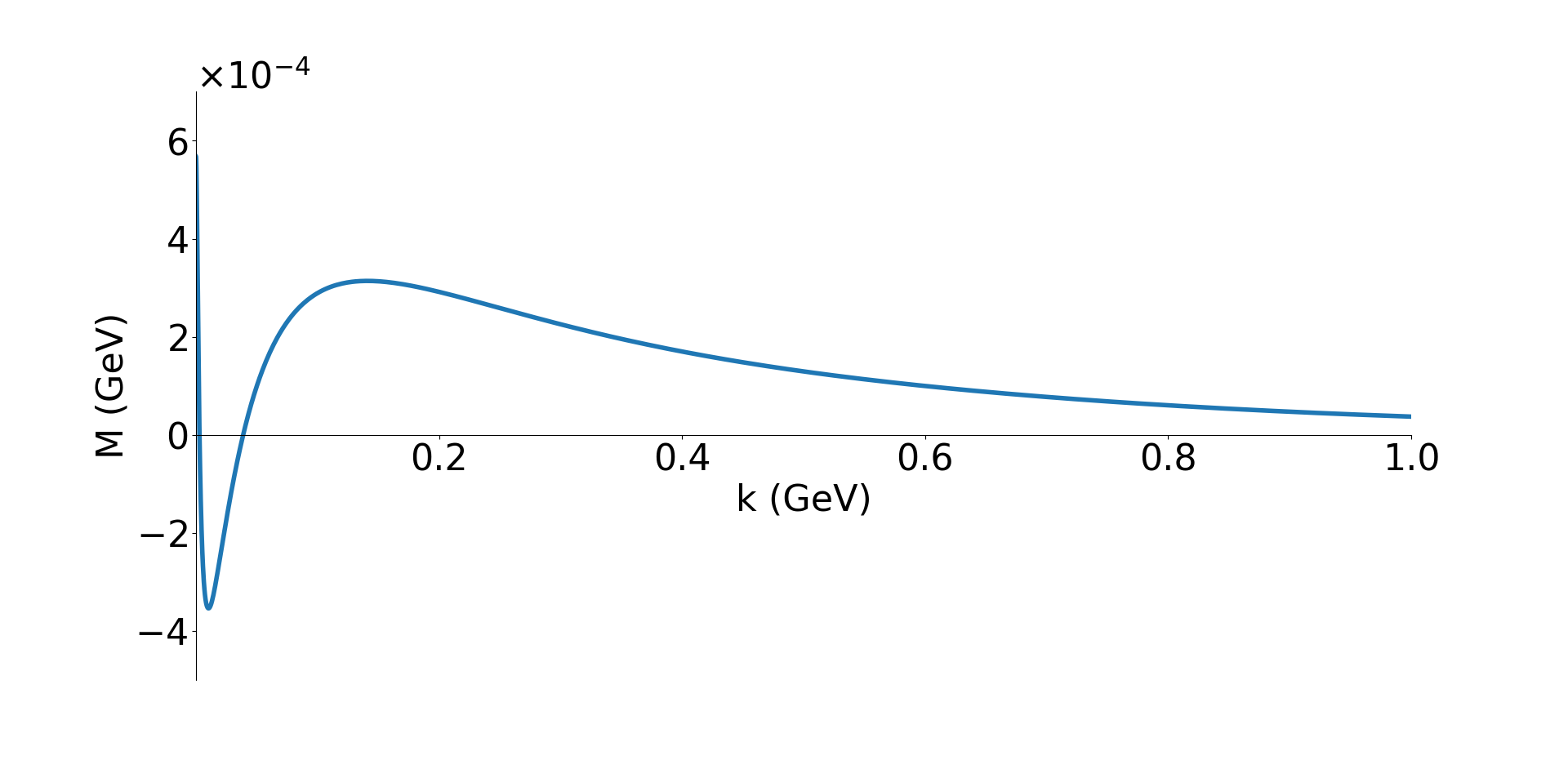}
\caption{\label{fig:masas} Constituent mass function of the quark $M(k)$ in the chiral limit $m_q=0$. From top to bottom, they correspond to the $\phi_k^0$, $\phi_k^1$ and $\phi_k^2$ solutions, here chosen to have $M(0)>0$.}
\end{figure}

It is straightforward to calculate the associated mass functions $M(k)$ for the quasiparticles (the bare particles at large momentum 
have a slowly running mass $m_q$) and constituent masses $M(0)$.
The relation between this constituent mass function and the gap angle is
\begin{equation}
    \sin \phi(k)=\dfrac{M(k)}{E}    
\end{equation}
with $E=\sqrt{M^2(k)+k^2}$.  For the case $m_q=0$, the gap angles are presented in Fig.~\ref{fig:sols0} and the dressed masses $M(k)$  in Fig.~\ref{fig:masas}. 
The solutions have an increasing number of nodes with the ground state BCS state having none, the first replica having exactly one zero,
and the second one having precisely two. Some version of a nonlinear Sturm-Liouville theorem must be at play.

\begin{figure}[t]
\includegraphics[scale=0.175,trim={0 1.5cm 0 2cm},clip]{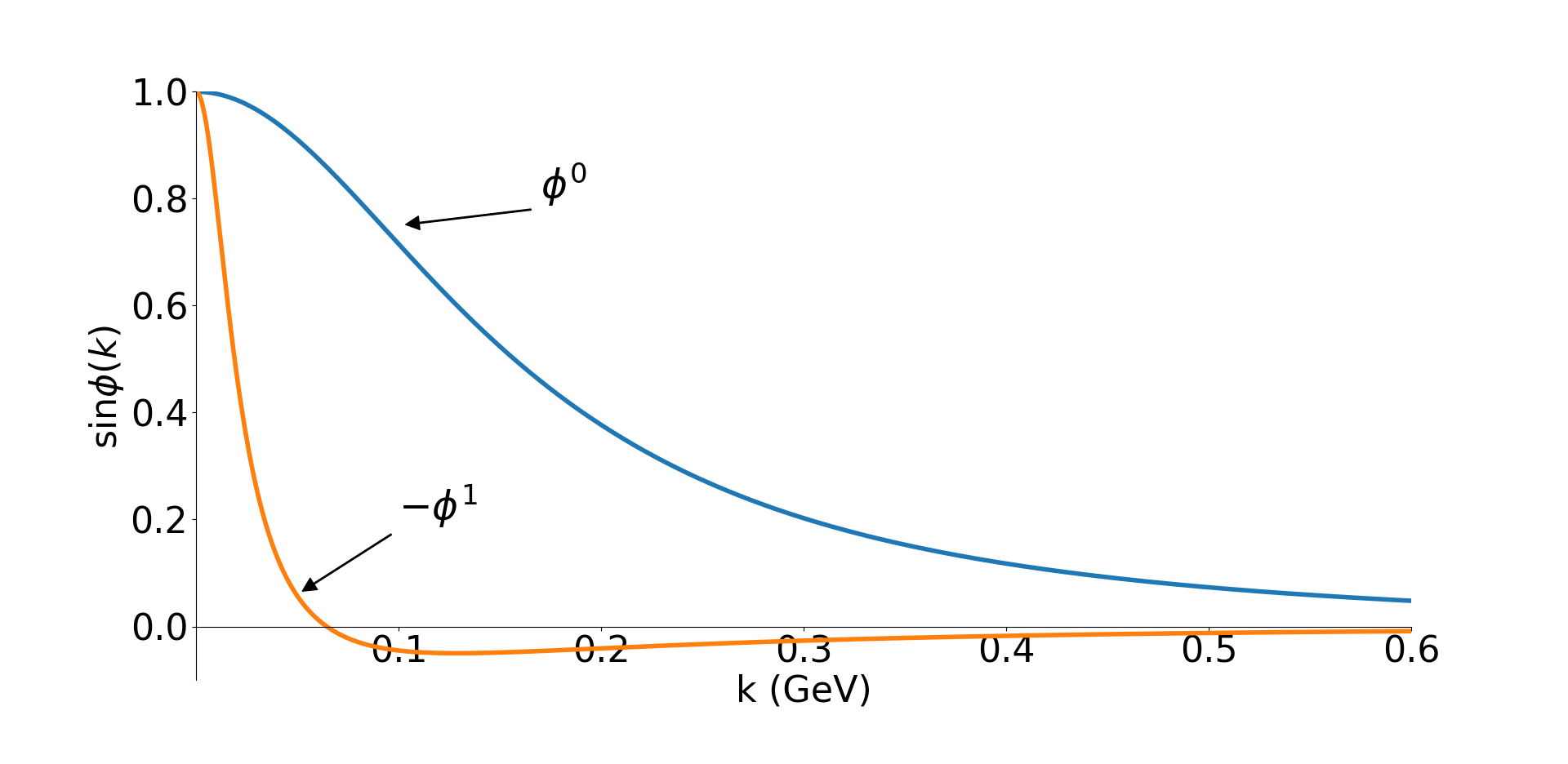}% Here is how to import EPS art
\caption{\label{fig:sols1} The first two (non-trivial) solutions of the mass gap equation with $m_q= 1 \text{ MeV}$. We show $\phi_k^1$ with a sign change to compare it with the chiral limit in Fig.~\ref{fig:sols0}.}
\end{figure}

\begin{figure}[t]
\includegraphics[scale=0.175,trim={0 0 0 1cm},clip]{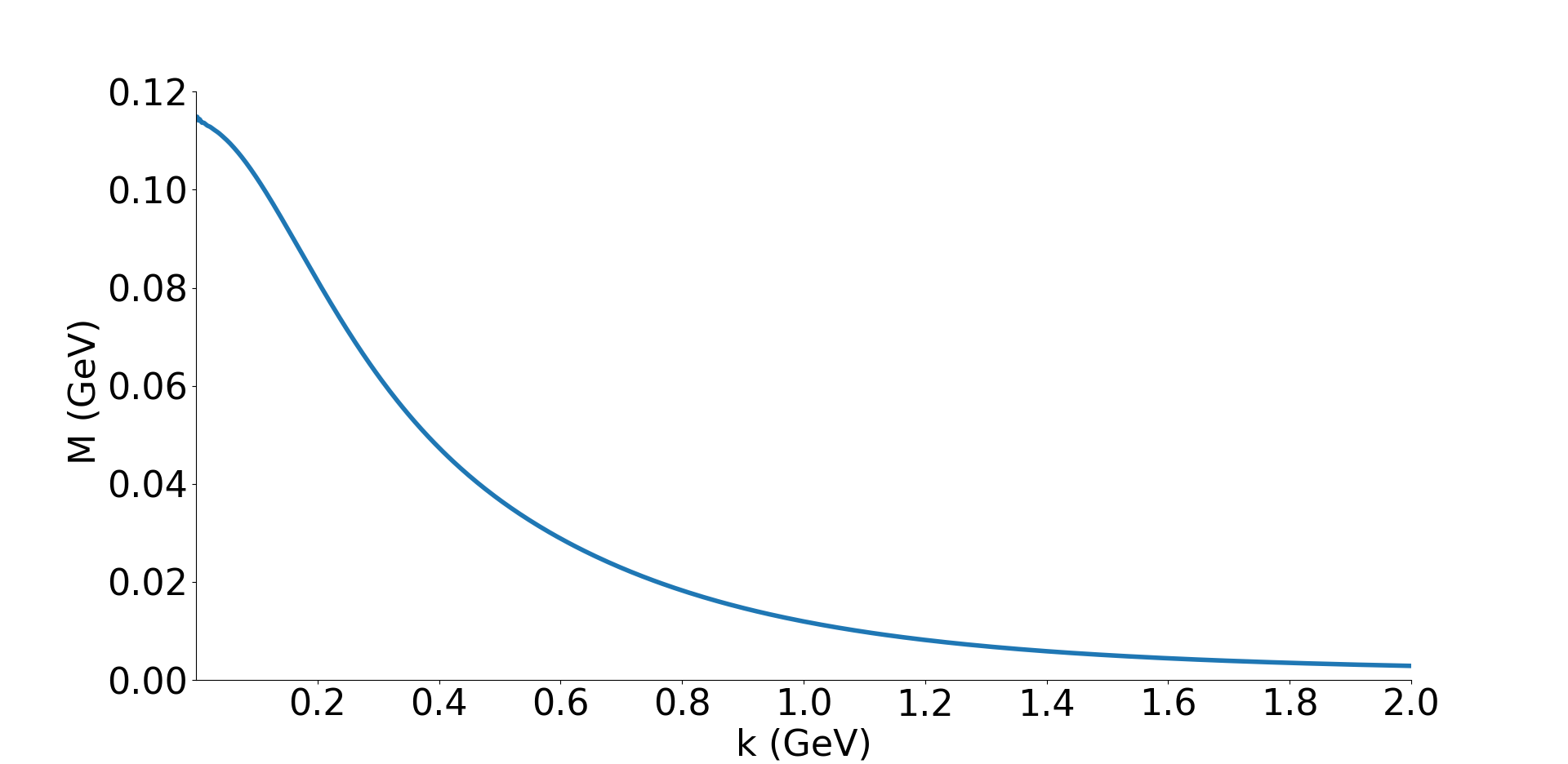}
\includegraphics[scale=0.175,trim={0 1cm 0 1cm},clip]{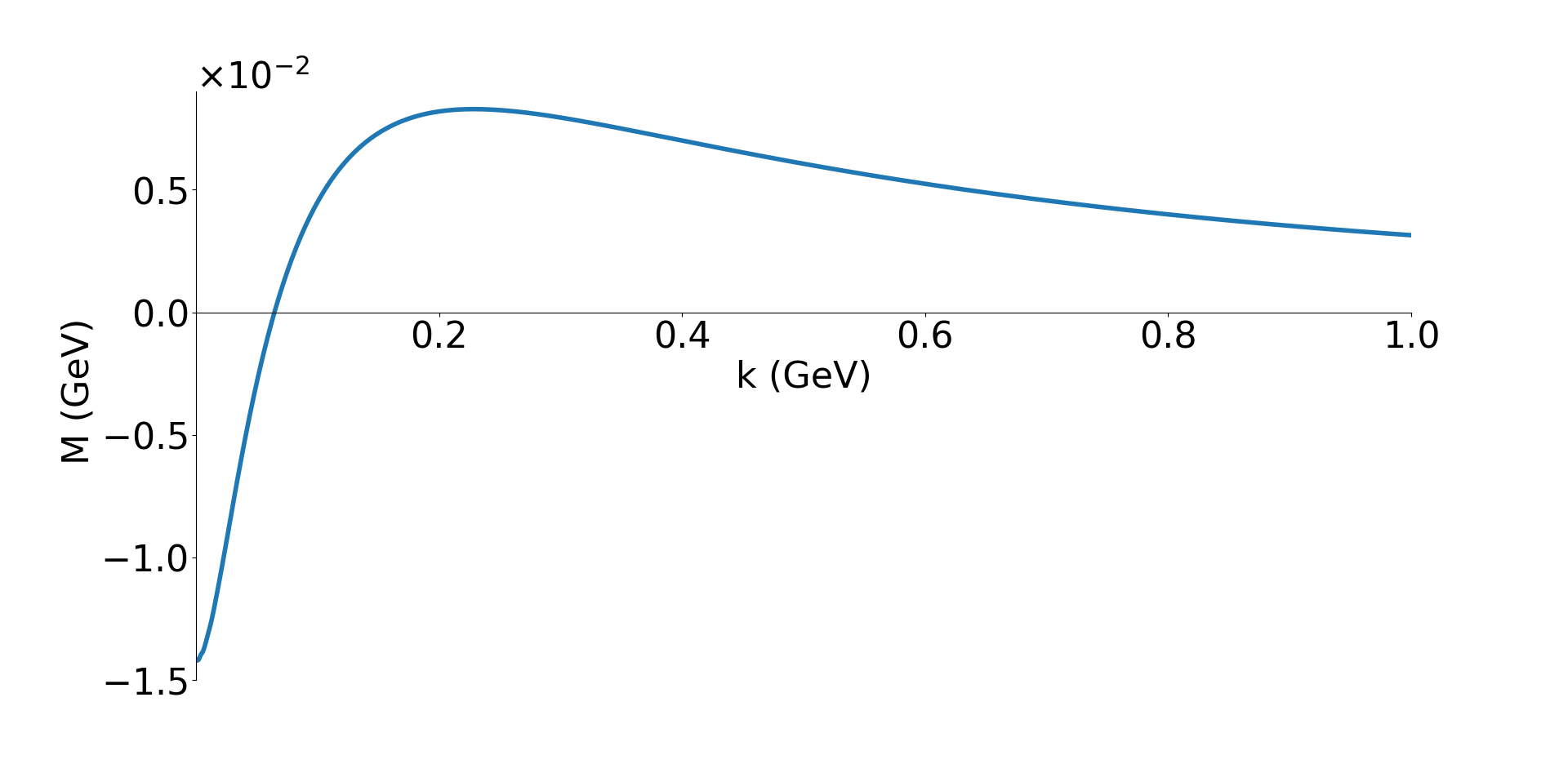}% 
\caption{\label{fig:masas2} Constituent mass function of the quark, $M(k)$, with $m_q= 1 \text{ MeV}$. Upper plot: $\phi_k^0$, lower plot:  $\phi_k^1$.}
\end{figure}

The addition of a small current quark mass $m_q=1 \text{ MeV}$ collapses the second replica and we only have
two solutions to the interacting gap equation, instead of three (Fig.~\ref{fig:sols1} and \ref{fig:masas2}): the BCS solution $\phi^0_k$ and the first replica $\phi^1_k$. Although we thoroughly examined the solution space, we weren't able to reproduce the second replica that we find in the chiral limit, which seems to be a critical point for that solution.

\begin{figure}[b!]
\centering
\includegraphics[scale=0.175,trim={0 0 0 2cm},clip]{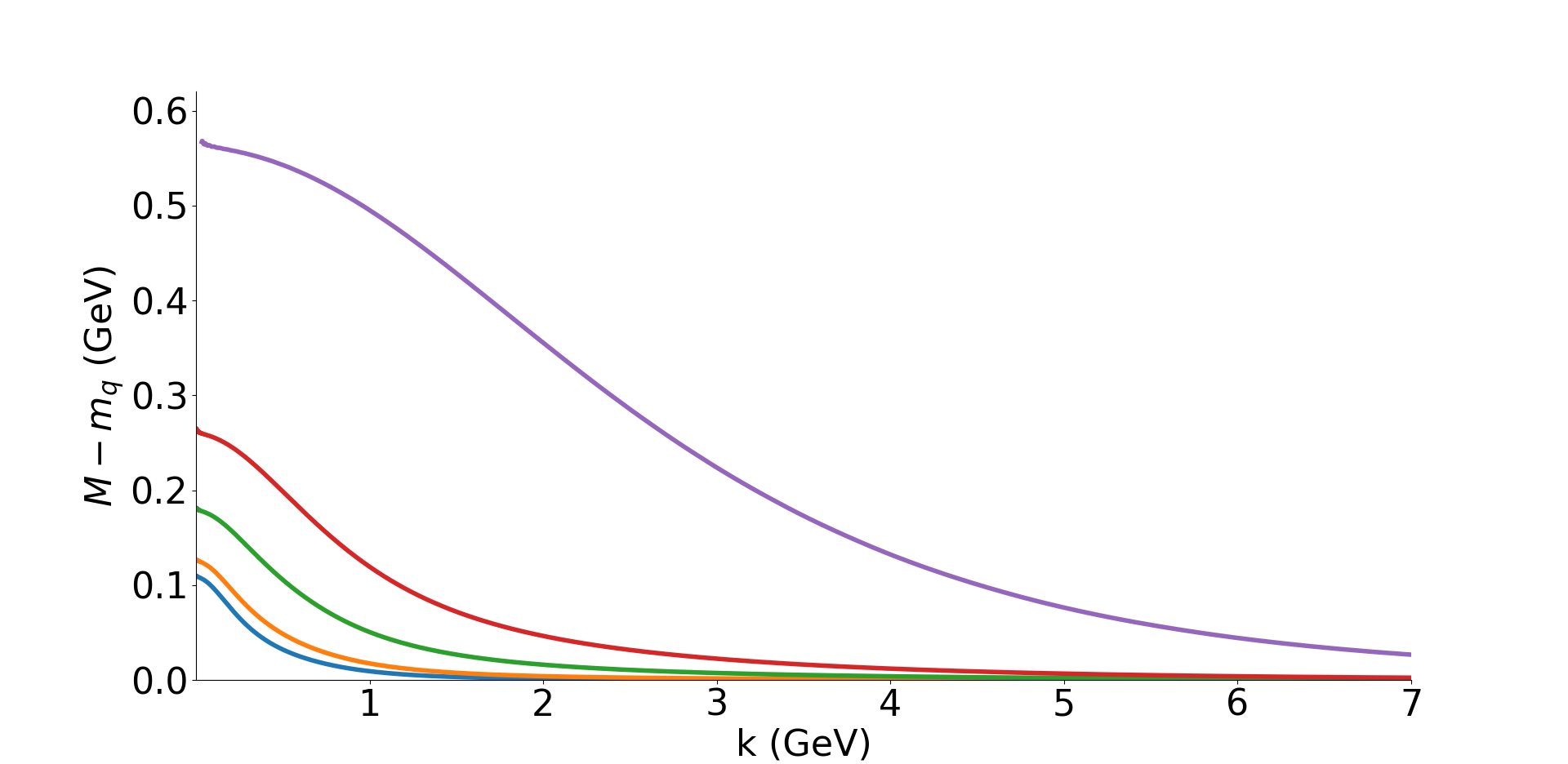}
\caption{\label{fig:rotura} Constituent quark mass function of the lowest BCS solution $\phi_k^0$ for different current quark masses. From bottom to top,  $m_q=0, 5, 30, 100, 1500 \text{ MeV}$. 
We plot $M(k)-m_q$ as the subtraction is convenient to compare the intensity of chiral symmetry breaking.}
\end{figure}

The constituent quark mass $M(k)$ characterizes the degree of symmetry breaking. We can analyze how the increase of the current quark mass $m_q$ affects the chiral symmetry breaking. We plot the dressed quark mass $M(k)$ in the BCS vacuum $\phi_k^0$ for different values of $m_q$ (Fig.~\ref{fig:rotura}). We can see for a small value of $m_q=5 \text{ MeV}$, which is characteristic of the $u$ and $d$ quarks, the degree of chiral symmetry breaking remains close to the chiral limit $m_q=0$. When we increase the current quark mass to $m_q=30,100 \text{ MeV}$, characteristic of the $s$ quark, the degree of chiral symmetry breaking has rapidly increased and when we arrive to masses of about the $c$ quark, with $m_q=1500 \text{ MeV}$, the symmetry is totally absent at low momentum, as seen in Fig.~\ref{fig:rotura}.

\newpage

%%%%%%%%%%%%%%%%%%%%%%%%%%%%%%%%%%%%%%%%%%%%%%%%%%%%%%%%%%%%%%%
\section{The replicae as saddle points of the energy density \label{sec:level4} }
%%%%%%%%%%%%%%%%%%%%%%%%%%%%%%%%%%%%%%%%%%%%%%%%%%%%%%%%%%%%%%%

The stability, metastability or instability of the replicae is
an interesting point that has been previously discussed in the literature. 
Once the gap function has been fixed, the finding of~\cite{Bicudo:2022hlm} with which we concur (Sec.~\ref{sec:mesons} below) is that the spectrum built thereover has no tachyons, with replicated hadrons all having nonnegative, and certainly real, masses.

What we instead examine in this section is what happens at the level of the gap function itself, whether the vacuum configuration can deform to another
one via a collective state, and not a few-body quasiparticle excitation. 

To be specific, we frame our calculation in the chiral limit $m_q=0$, since a small quark mass does not change the overall picture for the first replica, but it does not allow to study the second.  

The mass gap equation solutions are extremal points of the energy density  $\delta \langle \Omega| H |\Omega\rangle/\delta \phi_i(k)=0$, and
to study the stability of each solution we need to calculate the second functional derivative of the vacuum energy density
\begin{eqnarray}
     F(k,q)=\dfrac{\delta^2 \langle \Omega| H |\Omega\rangle}{\delta \phi_q \delta\phi_k}\ .
\end{eqnarray}
The fastest way to asses the positivity (or otherwise) of such a quadratic form is to look at the sign of the eigenvalues of its matrix evaluated at each solution $\phi_k^i$ of Eq.~(\ref{eq:massgap}). If all the eigenvalues are strictly positive, i.e. the matrix is definite positive, the solution corresponds to a minimum. If the eigenvalues are all negative, it is a maximum. Finally, if the eigenvalues are mixed, the solution is a saddle point. In the first case and interpreting $\phi_k$ as a classical effective field, there would not be any classical trajectory to decay from that solution to another and the vacuum would be metastable. However, in the last two cases there would be classical trajectories to decay and the vacuum would be unstable. The eigenvalue equation for this matrix reads
\begin{eqnarray}
  \int \dfrac{\dif^3 q}{(2\pi)^3} F(k,q) \psi_i(q)=\lambda_i \psi_i(k)
\end{eqnarray}
\hspace{0.5cm}
\noindent which in the model of Eq.~(\ref{eq:rho}) leads to
\begin{widetext}
\begin{equation}
\label{eq:eigen}
\begin{split}
   \lambda_i\psi_i(k)=& 6(kc_k+m_q s_k)\psi_i(k)-4\int \dif^3 q \left[\hat{V}(|\Vec{k}-\Vec{q}\,|)(s_k s_q+c_qc_kx)+2\hat{U}(|\Vec{k}-\Vec{q}\,|)s_k s_q+2c_kc_q\hat{W}(|\Vec{k}-\Vec{q}\,|)\right]\psi_i(k)\\
   &+4\int \dfrac{\dif^3 q}{(2\pi)^3}\left[\hat{V}(|\Vec{k}-\Vec{q}\,|)(c_k c_q+s_q s_kx)+2\hat{U}(|\Vec{k}-\Vec{q}\,|)c_k c_q+2s_k s_q\hat{W}(|\Vec{k}-\Vec{q}\,|)\right]\psi_i(q).
\end{split}
\end{equation}
\end{widetext}

\begin{table}[h]
\caption{\label{tab:eigen}
Summary of the nature of the mass gap equation solutions (``vacua''). Eq.~(\ref{eq:eigen}) is discretized over a radial momentum grid of 600 points to yield mostly positive eigenvalues, but we list here whether any of them is negative.}
\begin{ruledtabular}
\begin{tabular}{ccc}
\mbox{Vacuum}&\mbox{\mbox{Critical point}}&
\mbox{\# negative eigenvalues}\\
\hline
\mbox{BCS}&\mbox{Minimum}&0\\
\mbox{1st replica}&\mbox{Saddle point}& 1\\
\mbox{2nd replica}&\mbox{Saddle point}& 2\\
\mbox{\mbox{Perturbative}}&\mbox{Saddle point}&3\\
\end{tabular}
\end{ruledtabular}
\end{table}
We solve the eigenvalue problem numerically, finding the results summarised in Table~\ref{tab:eigen}. As we expected, the BCS vacuum $\phi^0_k$ is a minimum, so it is stable and thereon we usually model the QCD hadronic spectrum. The perturbative vacuum is a saddle point and therefore, unstable. The fact that it has exactly three negative eigenvalues (the paths pointing towards each of the solutions of the interacting gap equation, the vacuum and the two replicae that have negative energy density) is very suggestive that we have found
all negative-energy solutions to the nonlinear gap equation in the interacting theory, for which we have no known theorem specifying the number and nature of such solutions.

Although the traditional picture of the Mexican hat suggests the idea that the perturbative vacuum has to be a maximum, that picture is just a graphical assistance to visualize the idea of the two classes of vacua. 
Actually, we are working in the space function $L^2$ for the gap function $\sin \phi(k)$, which is an infinite dimensional space and the vacuum picture is not as neat as a Mexican hat. However, the instability of the perturbative vacuum triggers spontaneous symmetry breaking of the theory and its (classical) decay to the BCS vacuum. Finally, we find both replicae $\phi^1_k$ and $\phi^2_k$ are also saddle points and hence they seem to be classically unstable to collective perturbations that dislocate the entire vacuum function $\phi(k)$ by creating an infinite number of quark-antiquark pairs with specific weights in the directions of the negative eigenvalues.

At a second, few--body level, examining the creation of meson-like $q\Bar{q}$ states over the replica,
 \begin{equation}
    \langle \Omega_i|[H,\int \dif^3 k\, \Psi B^\dagger_k D^\dagger_{-k} ]|\Omega_i\rangle
\end{equation}
is a matrix with only positive eigenvalues, where $|\Omega_i\rangle$ $i=1,2$ is the first replica and second respectively. This means all the meson excitations found over the replicae have positive mass squared (no tachyons). 
We postpone the numerical results for this spectrum until after we have further discussed the collective negative eigenvalues in the next Sec.~\ref{sec:slices}.

%%%%%%%%%%%%%%%%%%%%%%%%%%%%%%%%%%%%%%%%%%%%%%%%%%%%%%%%%%%%%%%
\section{From replica to replica slicing the function space \label{sec:slices}}
%%%%%%%%%%%%%%%%%%%%%%%%%%%%%%%%%%%%%%%%%%%%%%%%%%%%%%%%%%%%%%%

In this section we look at the relative energy density of the replicae, to ascertain their relative stability 
as determined by energetic considerations alone. 
 We fix the chiral limit $m_q=0$ where all solutions are active. 
 (A small quark mass entails a small increase in $M(k)$, with no significant effect on the general properties of the remaining vacuum states, but we lose one replica).
 
A visual way to represent the relative placement of the solutions to the gap equation is to slice the infinitely--dimensional function space 
with a one--parameter curve that passes through two or more of those solutions, interpolating between them.

\begin{figure*}
    \centering
    \includegraphics[width= 7.5cm]{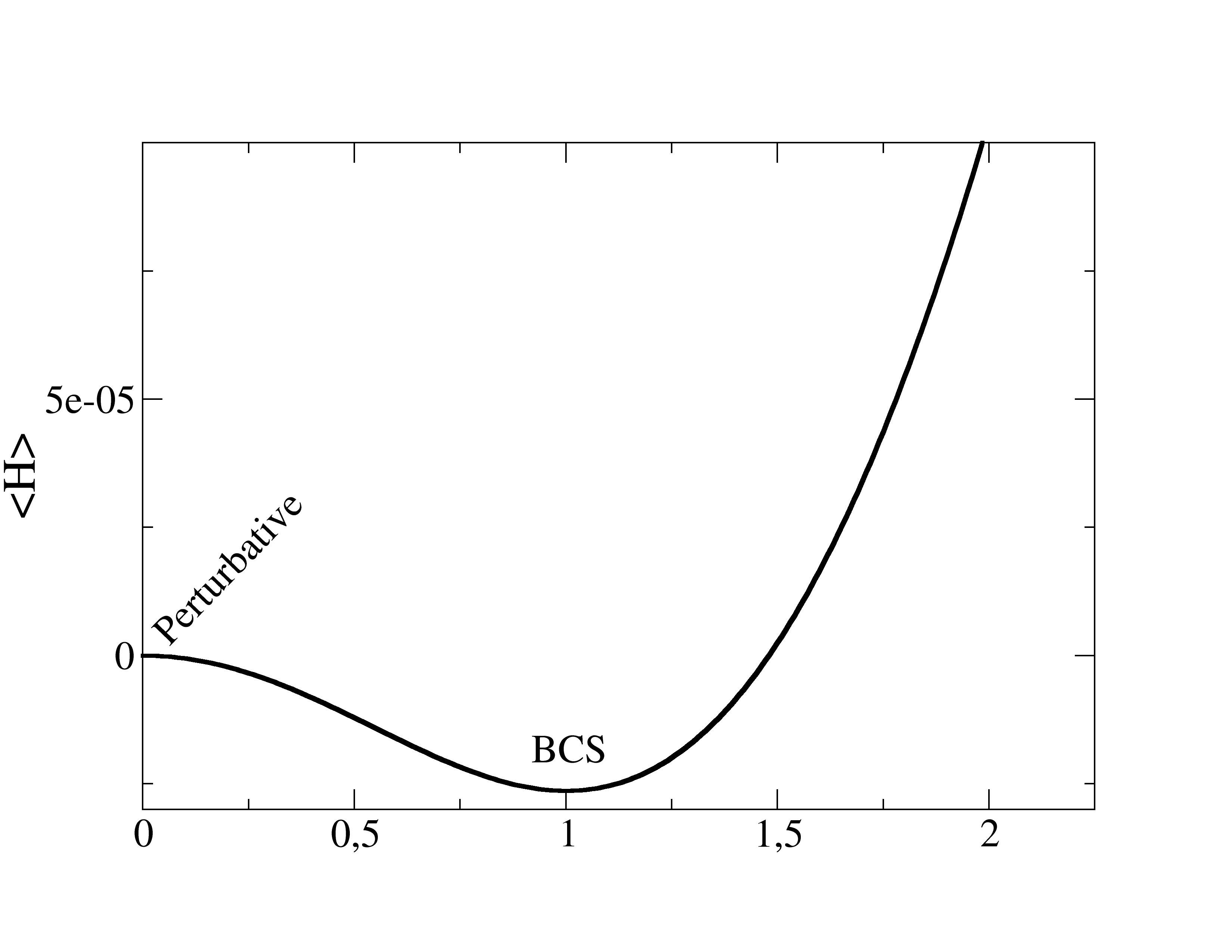}
    \includegraphics[width = 7.5cm]{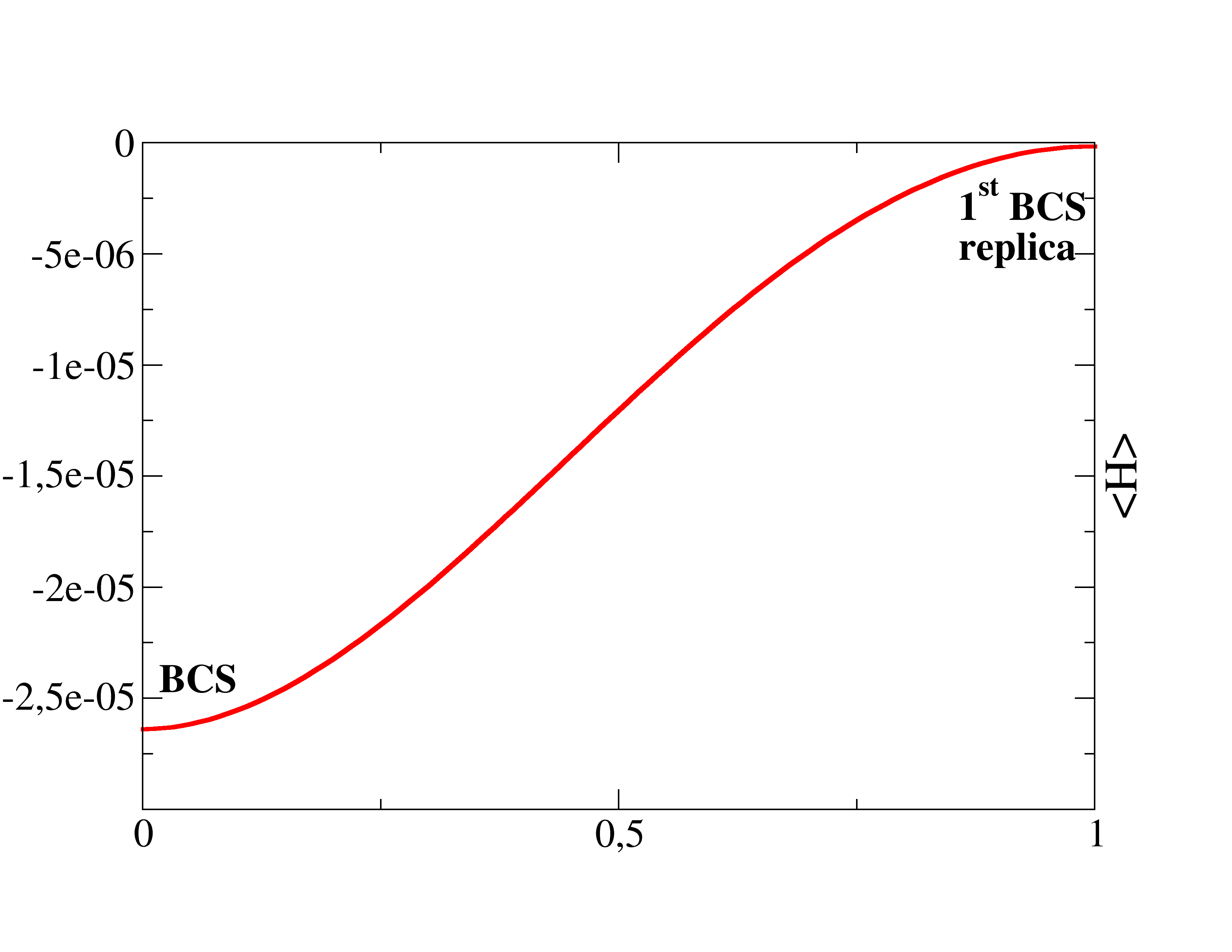}
        \includegraphics[width = 7.5cm]{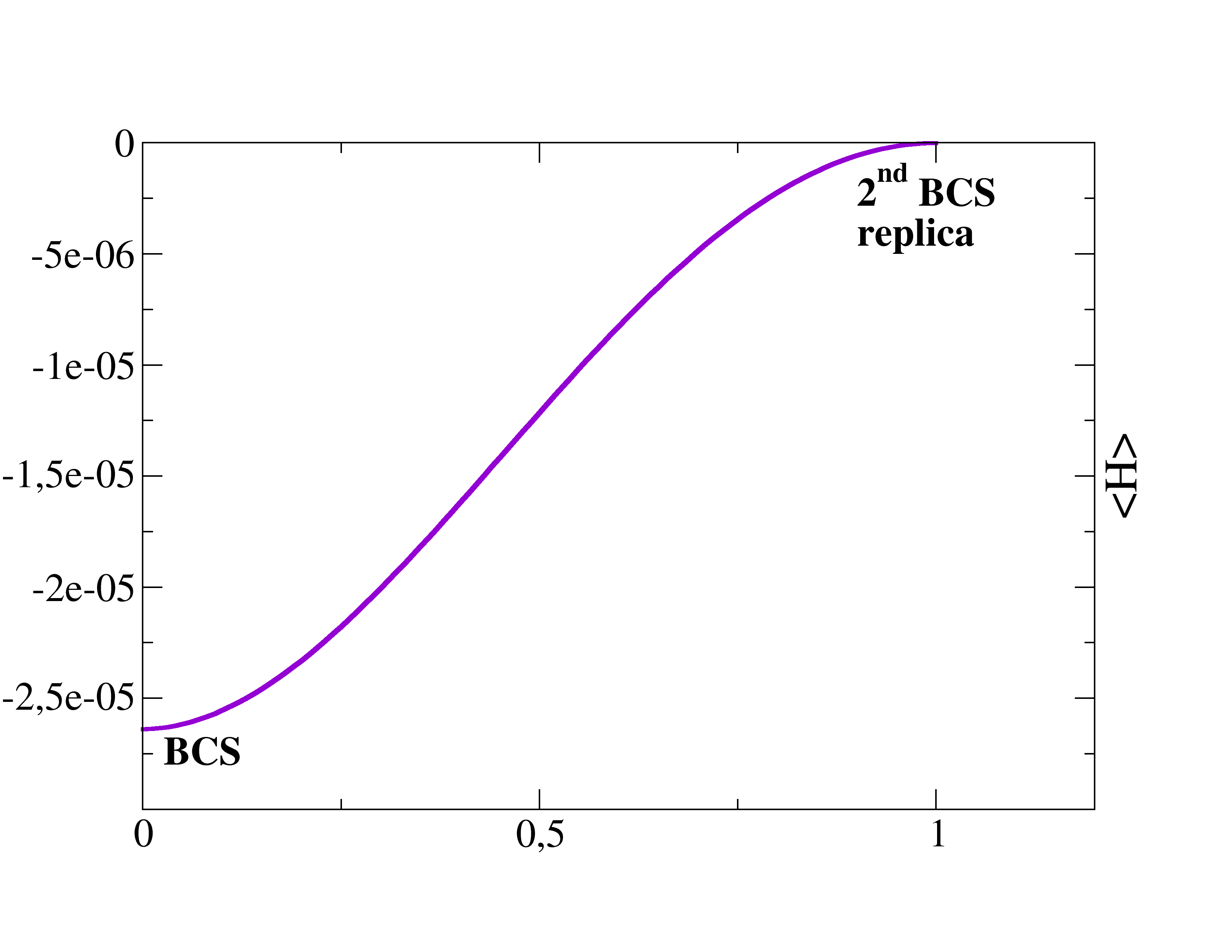}
    \includegraphics[width = 7.5cm]{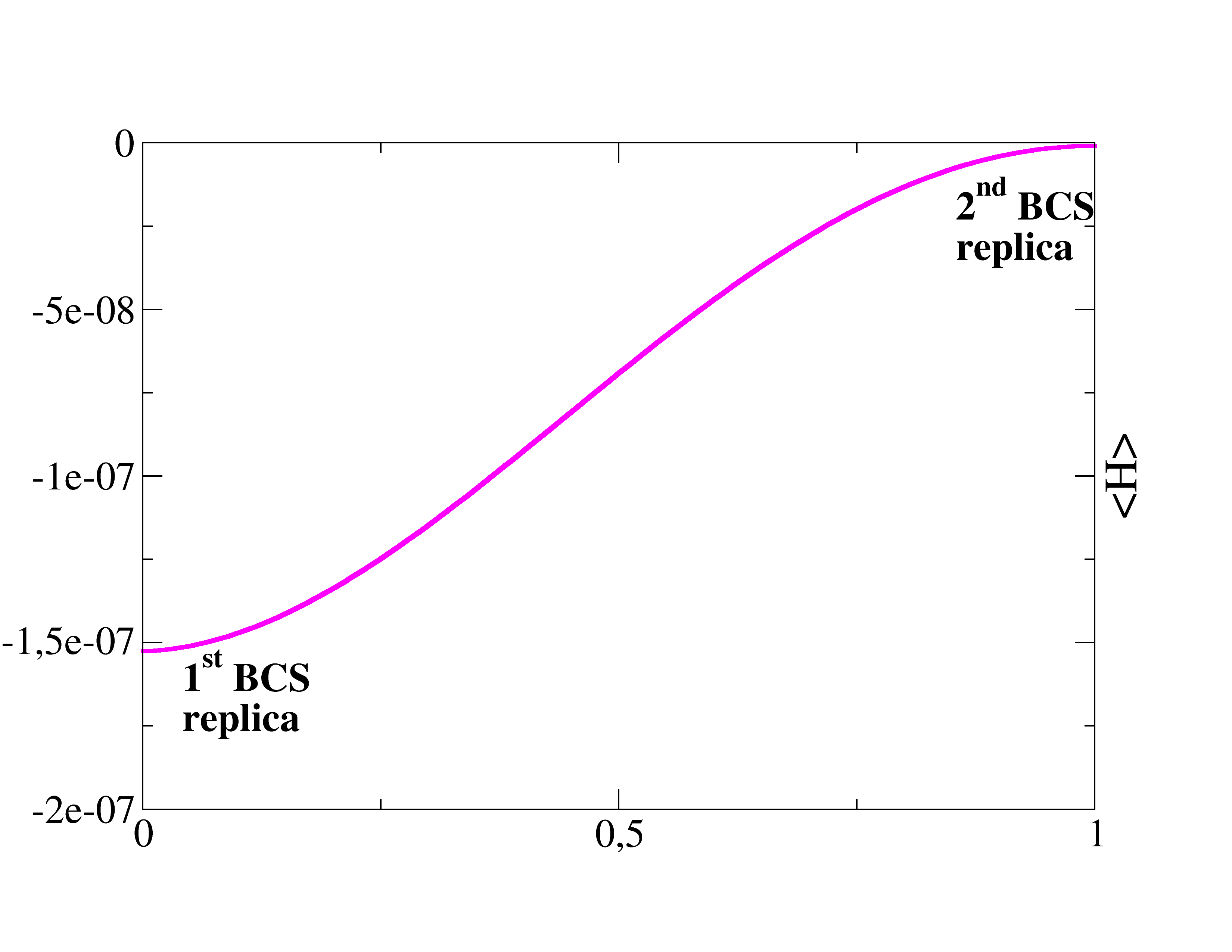}
    \includegraphics[width = 7.5cm]{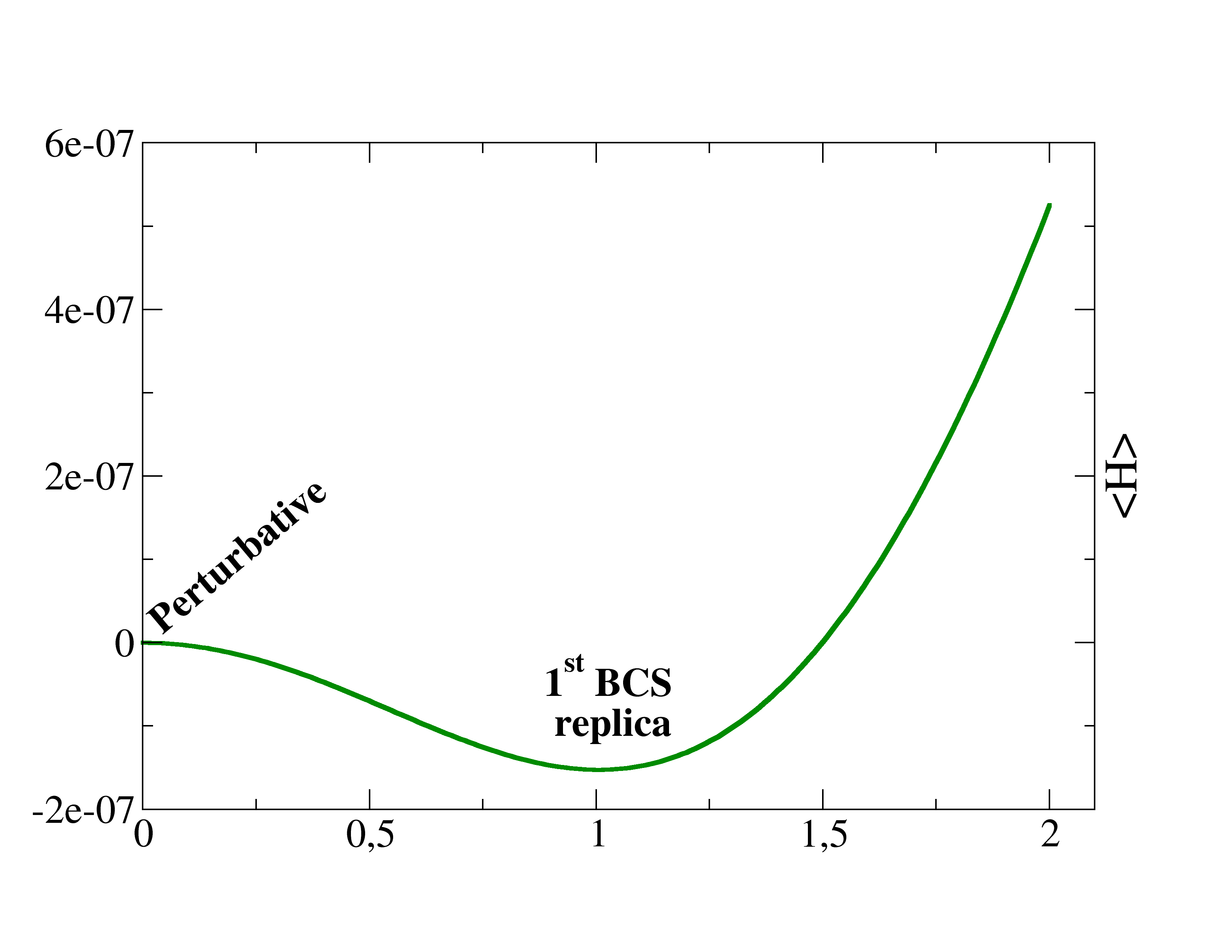}
    \includegraphics[width = 7.5cm]{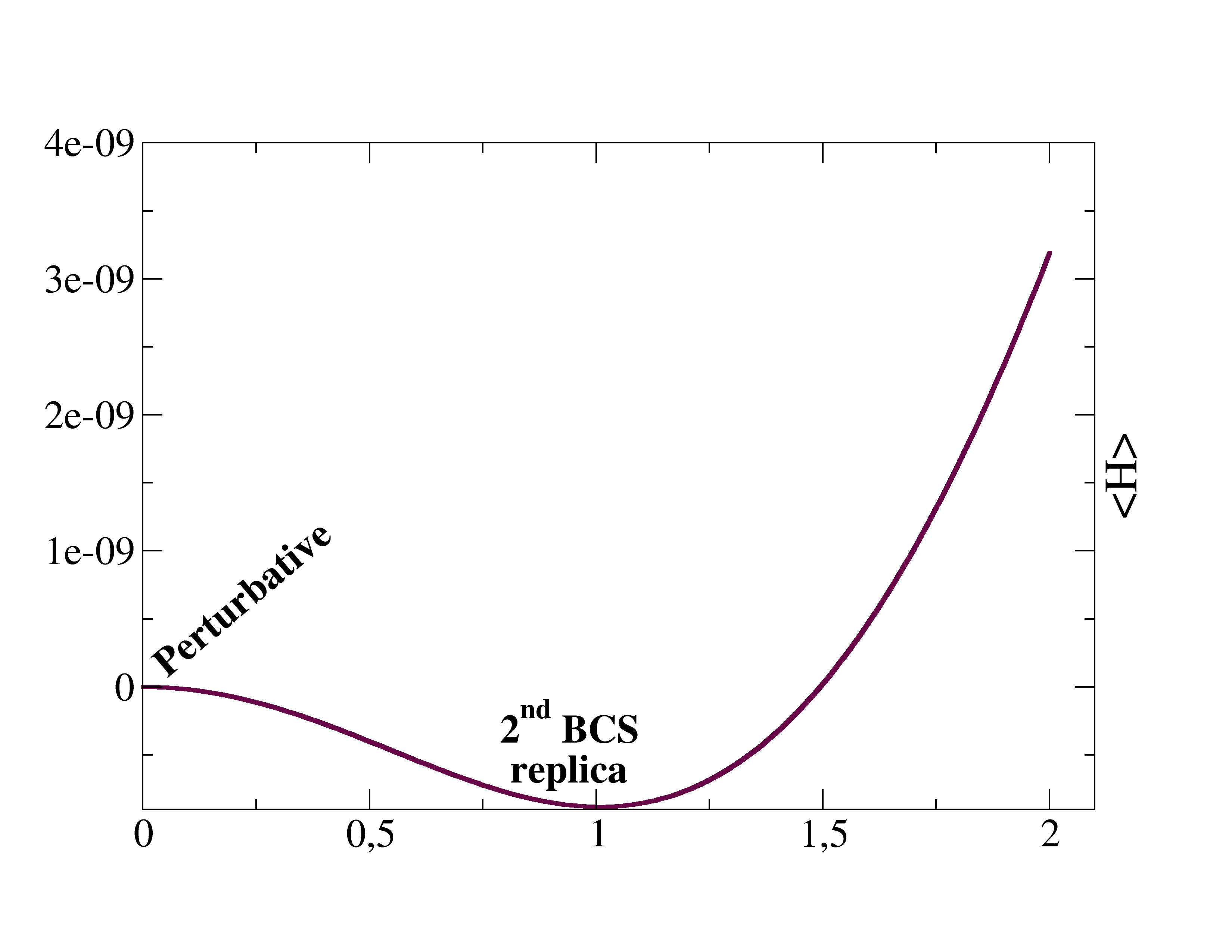}
\caption{\label{fig:detailedinterpols} Vacuum energy density (in GeV$^4$) as function of a parameter $\alpha$ that linearly interpolates
between pairs of replica states, $\varepsilon(\alpha) = \langle H \rangle (\alpha)$. We see that the commonly used BCS ground state has indeed the lowest energy, and that the two replicae lie between this and the perturbative vacuum (ground state of the free Hamiltonian).}
\end{figure*}

A detailed appraisal can be obtained by a simple linear interpolation between pairs of solutions, 
\begin{equation}
\phi(k) = \alpha \phi_i(k) + (1-\alpha) \phi_j(k)\  \ \ \ i\neq j\ . 
\end{equation}
The result of the exercise is shown in figure~\ref{fig:detailedinterpols}, that plots $\langle H \rangle (\alpha)$, the energy density along that path.
The characteristic depth of the ground-state BCS solution with respect to the perturbative vacuum is, as seen in the figure, $2.5\times 10^5$ GeV$^4\simeq \frac{1}{4}$ (100 MeV)$^4$, with the two replicae much closer to the zero level of the perturbative $|0\rangle$.

We can see therefrom that there is a deepest or ground state (simply denoted as BCS), two replicae with higher energy, and the perturbative ground state that minimizes the free Hamiltonian, which has the highest energy density of the four. 
Most interestingly, there is no potential barrier separating the various states: in a finite volume, the system can roll down from configuration to configuration until the minimum is reached.

We can depict all states together if we employ a more complicated interpolating function that slices through function space 
stopping at each of the replicae.  A polynomial function that can serve this purpose is
\begin{equation} \label{interpol2}
f(\alpha)=A\alpha^3+B\alpha^2+C\alpha+D    
\end{equation}
where $A,\ B,\ C,\ D$ are combinations of the solutions $\phi_i$ so as to satisfy the following conditions for the interpolation to pass 
by each of the relevant states under study.
First, $f(\alpha=0) = \arrowvert 0\rangle$ starts at the trivial vacuum; then,  when $\alpha=1/2$ $f$ is the BCS vacuum; for $\alpha=1$ 
we reach the first replica and when $\alpha=3/2$ the second one. 

A convenient choice is then
\begin{equation} \label{interpol2infull}
\begin{split}
    f(\alpha)=&\left[\dfrac{4}{3}(\phi^2_k-\phi^{\rm pert}_k)-4(\phi^1_k-\phi^{\rm pert}_k)+4(\phi^0_k-\phi^{\rm pert}_k)\right]\alpha^3\\
    +&\left[-2(\phi^2_k-\phi^{\rm pert}_k)+8(\phi^1_k-\phi^{\rm pert}_k)-10(\phi^0_k-\phi^{\rm pert}_k)\right]\alpha^2\\
    +&\left[\dfrac{2}{3}(\phi^2_k-\phi^{\rm pert}_k)-3(\phi^1_k-\phi^{\rm pert}_k)+6(\phi^0_k-\phi^{\rm pert}_k)\right]\alpha\\
    +&\phi^{\rm pert}_k
\end{split}
\end{equation}

\begin{figure}
    \centering
 \includegraphics[width= \columnwidth]{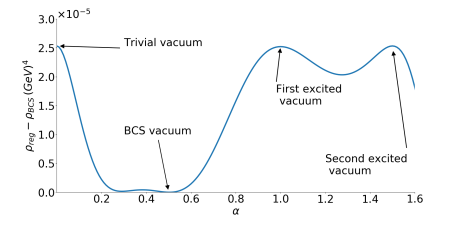}
\includegraphics[width= \columnwidth]{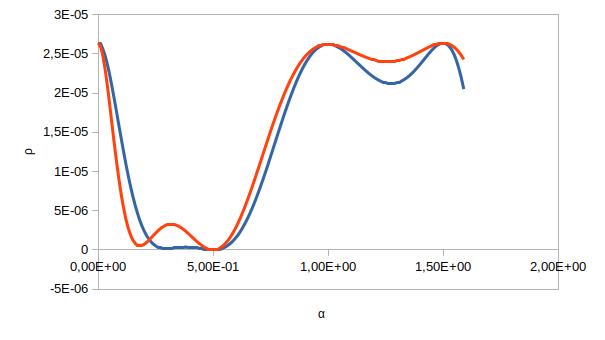}
    \caption{Polynomial interpolations between the solutions to the mass gap equation found in the chiral limit.
    Upper plot: from Eq.~(\ref{interpol2}). Lower plot: we add the interpolation from Eq.~(\ref{interpol3}) for comparison. The excited
    replicae, which are saddle points of $\varepsilon$, can appear as maxima in such cuts of function space that link them with the ground state.}
    \label{fig:birdseyeinterpol}
\end{figure}

We have to numerically calculate the energy density as an integral over the interpolating $\phi$ function,
\begin{equation}
\begin{split}
     \rho &=-\dfrac{3}{\pi^2} \int_0^\infty \dif k\,(k^3 c_k+m_qk^2 s_k)\\
     &-\dfrac{1}{4\pi^4}\int_0^\infty \dif k\, k^2\int_0^\infty \dif q\, q^2 \big[ V_0(1-s_k s_q) -V_1 c_k c_q \\
    &-2U_0(1+s_k s_q) -2c_kc_qW_0\big]
\end{split}
\end{equation}
in $\phi(k)=[f(\alpha)](k)$ for $\alpha \in [0,3/2]$.  We subtract the trivial vacuum contribution $\rho_t=\langle 0| H |0\rangle/V$ to control the UV divergence and thus the three solutions to the interacting gap equation will take negative values of the energy density,
\begin{equation}
\begin{split}
    \rho_{reg}  &=-\dfrac{3}{\pi^2} \int_0^\infty \dif k\,(k^3 (c_k-1)+m_qk^2 s_k)\\
    &+\dfrac{1}{4\pi^4}\int_0^\infty \dif k\, k^2\int_0^\infty \dif q\, q^2 \big[ V_0s_k s_q +V_1 (c_k c_q-1) \\
    &+2U_0s_k s_q +2W_0(c_kc_q-1)\big].
\end{split}
\end{equation}
The numerical result is presented in Fig.~\ref{fig:birdseyeinterpol}.

For comparison, we add a second interpolation with the following form:
\begin{equation} \label{interpol3}
f(\alpha)=A\alpha^2+B\alpha+C+\frac{D}{\alpha+1}    
\end{equation}

The reader may note that in the figure, not all the solutions to the gap equation appear as minima.
They are all in fact extrema of the energy--density expectation value $\varepsilon$, but since the excited 
states appear as saddle points (which will be ascertained shortly) they can appear as local maxima depending on the sliver of function space taken.

In the upper plot of Fig.~\ref{fig:birdseyeinterpol} there appear other maxima/minima which do not correspond to any of the solutions $\phi_i$: $\varepsilon$ has non vanishing derivatives in other directions not shown. They are just an artifact of the parametrization.
This is why having a second one to compare (lower plot) is useful as it clarifies the position of the actual replicae.

 Since such replicae appear in these sin$\phi$-space cut as two local maxima, because the parametrization links them to the ground state 
 which has smaller energy   (so they do not need to appear as maxima in other cuts, as they are saddle points), 
 and the energy density is a smooth function, there naturally is an apparent minimum between them, but this is not a physically relevant state as it does not solve the gap equation (it is not a critical point in other directions in function space). 

 Having exhausted the discussion about the collective excitations modifying the gap function, we now proceed to the few-body excitations over each of the replicae.

%%%%%%%%%%%%%%%%%%%%%%%%%%%%%%%%%%%%%%%%%%%%%%%%%%%%%%%%%%%%%%%%%%%%%%%%
\section{Spectrum over the replica vacua \label{sec:mesons}}
%%%%%%%%%%%%%%%%%%%%%%%%%%%%%%%%%%%%%%%%%%%%%%%%%%%%%%%%%%%%%%%%%%%%%%%%

\begin{figure}[b!]
\includegraphics[width=\columnwidth]{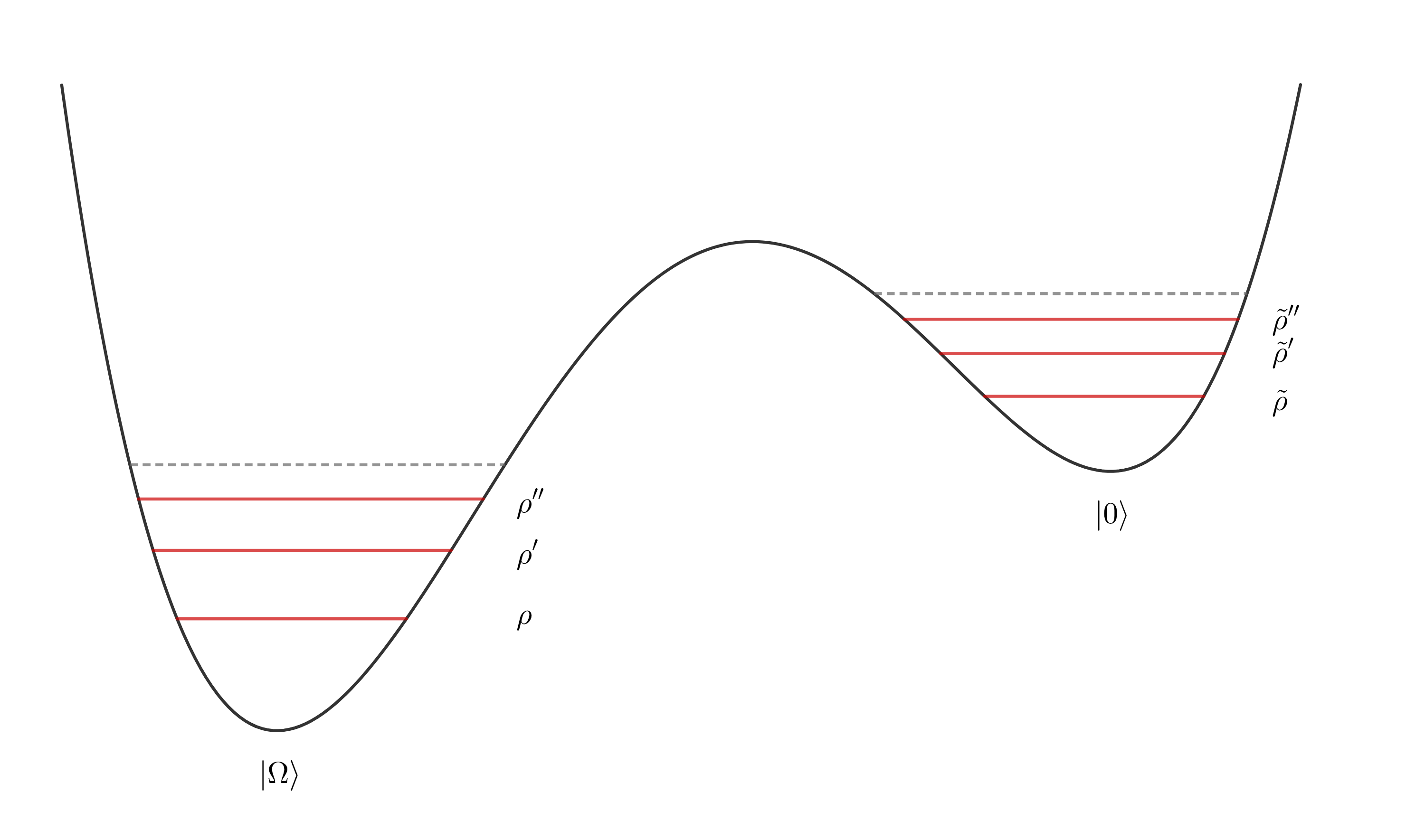}
\caption{\label{fig:parallelspectra}  
We sketch the spectrum of vector-meson excitations as energy levels above both the ground-state BCS vacuum $|\Omega\rangle$ and the perturbative vacuum $|0\rangle$: the two ``parallel'' spectra are quite similar as visible in Table~\ref{table:Erho} below.
}
\end{figure}

We now address the excitation spectrum of mesons on top of the ground state and of the replicae, producing ``replicated hadrons'' that differ from ordinary ones in the underlying BCS ground state. The situation is visually explained in figure~\ref{fig:parallelspectra}, that shows two similar sets of hadrons excited over two different vacua.

We employ the RPA (Random Phase Approximation), that is 
the instantaneous generalization of the Schr\"odinger equation to a degree of freedom with forward and backward propagation components, used in nuclear physics to properly describe the zero mode associated with translations of the nucleus as a whole,
and in this hadron physics field to properly implement chiral symmetry breaking for the Goldstone bosons, the pions~\cite{Bicudo:1989si}. 

Omitting the color index, the pion creation operator is
\begin{eqnarray}
    Q^\dagger(RPA)=\sum _{\alpha\beta}\int \frac{d\vec{k}}{(2\pi)^3} \times \nonumber \\
    \left[\mathcal{X}_{\alpha\beta }(\vec{k)}B^\dagger_\alpha(\vec{k}) D^\dagger_\beta(-\vec{k})-\mathcal{Y}_{\alpha\beta}(\vec{k})B_\alpha(\vec{k}) D_\beta(-\vec{k})\right]
\end{eqnarray}
featuring two wavefunctions $\mathcal{X}$ and $\mathcal{Y}$ 
\begin{eqnarray}
    \mathcal{X}(\vec{k})=\frac{1}{\sqrt{4\pi}}\frac{i(\sigma_2)_{\alpha\beta}}{\sqrt{2}} \frac{\delta_{ij}}{\sqrt{3}}X^\nu (k)\nonumber \\
    \mathcal{Y}(\vec{k})=\frac{1}{\sqrt{4\pi}}\frac{i(\sigma_2)_{\alpha\beta}}{\sqrt{2}} \frac{\delta_{ij}}{\sqrt{3}}Y^\nu (k)\ .
\end{eqnarray}
The angular wavefunction for an s-wave meson is simply $Y^0_0(\hat{k})=\frac{1}{\sqrt{4\pi}}$, 
the spin wavefunction $i\sigma_2/\sqrt{2}$ coincides with the Clebsch-Gordan coefficient yielding $S=0$, and the color one $\delta_{ij}/\sqrt{3}$ is simply a normalized singlet. This projects the eigenvalue equation over the pion's $J^{PG}=0^{-+}$ quantum numbers. The radial wavefunctions for the forward and backward propagating components are denoted as
 $X^\nu$ and $Y^\nu$, respectively. 
The quasiparticle annihilation operator $Q$ satisfies $Q\arrowvert {\rm RPA}\rangle=0$ for an unspecified RPA ground state that
is more correlated than the BCS one.

The meson mass can be obtained as the eigenvalue of the coupled system of equations for $ \mathcal{X}$ and $\mathcal{Y}$,
\begin{eqnarray}
    \langle{RPA}|Q_{\rm RPA}\big [H,B^\dagger_\alpha(\vec{k}) D^\dagger_\beta(-\vec{k})\big]|{RPA}\rangle  &=& M_\pi^\nu \mathcal{X}(\vec{k}) \nonumber \\
    \langle{RPA}|Q_{\rm RPA}\big [H,B_\alpha(\vec{k}) D_\beta(-\vec{k})\big]|{RPA}\rangle
     &=&-M_\pi^\nu \mathcal{Y}(\vec{k}) \ .\nonumber \\
\end{eqnarray}

These are reduced to a linear system, under the usual bosonization approximation (in practice, setting the RPA vacuum to be the BCS 
vacuum once the commutators have been computed),
\begin{eqnarray}
   \int^\infty_0\frac{dq}{6\pi^2}q^2\big[K(k,q)X^\nu(q)+K'(k,q)Y^\nu(q)\big]\nonumber \\
    + 2\epsilon X^\nu(k) =M_\pi^\nu X^\nu (\vec{k})
\nonumber \\
   \int^\infty_0\frac{dq}{6\pi^2}q^2\big[K(k,q)X^\nu(q)+K'(k,q)Y^\nu(q)\big]
    \nonumber \\ + 2\epsilon Y^\nu(k)+=  -M_\pi^\nu Y^\nu (\vec{k}) \label{RPApions}
\end{eqnarray} 
featuring the kernels~\cite{Llanes-Estrada:2004edu}
\begin{eqnarray}
    K(k,q)=(1+s_kc_q)V_0+2(1-s_ks_q)U_0+c_kc_q(V_1-2W_0)\nonumber \\
     K'(k,q)=(1-s_kc_q)V_0+2(1+s_ks_q)U_0-c_kc_q(V_1-2W_0) \nonumber \\
\end{eqnarray}
and also the quark/antiquark self-energy obtained from the one-body piece of the Hamiltonian
\begin{equation}
    \epsilon_k=m_qs_k-kc_k-\int ^\infty_0\frac{d\vec{p}}{6\pi^2}[s_ks_p(V_0+2U_0)+c_kc_p(V_1+2W_0)]\ .
\end{equation}
By itself this self-energy is infrared divergent, but in combination with the two-body pieces, it yields a finite 
two-body RPA system due to cancellations induced by a Ward identity, due to the model's global color symmetry.

These two radial-momentum equations are numerically solved with an integrator and an eigensolver. 
Each eigenvalue corresponds to a meson with the quantum numbers of the pion built over the vacuum state of each of 
the replicae.

In table~\ref{Table:firstspectrum} we present the energy eigenvalues (in MeV) obtained in the RPA approximation, for mesons over the ground state and also over the replicae, and finally, over the perturbative vacuum, all for $m_q=0$.

\begin{table}[h]
\centering
\caption{Experimental meson spectrum from~\cite{ParticleDataGroup:2018ovx} (last row) and theoretical model computations over the different ground states with pseudoscalar quantum numbers  $0^{-+}$. All masses in MeV, are computed in the chiral limit $m_q=0$ so that the pion is a strict Goldstone boson over any vacuum that spontaneously breaks chiral symmetry.  \label{Table:firstspectrum}}
\begin{tabular}{|c|c|c|c|c|c|} \hline
\textbf{Vacuum} & $\pi$ & $\pi(1300)$ & $\pi(1800)$ & -- &  -- \\
\hline
$|\Omega\rangle$    & $\simeq 0$ & 1278          & 1968        & 2513     & 2967   \\ \hline
$|\Omega'\rangle$   & $\simeq 0$ & 1302          & 1997        & 2542     & 2995   \\ \hline
$|\Omega''\rangle$  & 0.4        & 1304          & 1999        & 2544     & 2997  \\ \hline
$|0\rangle $        & 500        & 1486          & 2018        & 2136     & 2656   \\ \hline
Experimental & 138      & 1300$\pm100$  & 1812$\pm12$ & --       & --     \\ \hline
\end{tabular}
\end{table}

Although the experimental pseudoscalar spectrum with light quarks has only three states up to 
$\pi(1800)$, we have quoted the next two eigenvalues yielding possible quark-antiquark mesons $\pi(2500)$ and $\pi(3000)$. 
One would expect those to be very broad as they have numerous open decay channels. 

The model yields quite accurately a massless pion, implementing Goldstone's theorem in the presence of chiral symmetry breaking. 
The pion mass is vanishing over the actual ground BCS state $\Omega$ as well as over its two replicae in the chiral limit (the 0.4 MeV value over the second replica is compatible with zero within the numerical error of the computation of order 1-2 MeV, whose reduction is unnecessary in view of the much larger systematic model uncertainties).

Over the perturbative vacuum, however, since it is chirally symmetric, the pion is not massless but rather it acquires some 500 MeV of mass corresponding to a light quark-antiquark excitation with reduced chromomagnetic energy (the spins are antialigned) nor centrifugal energy (the orbital angular momentum is $L=0$).

A fun fact is that, from this table with the pseudoscalar spectrum alone, it would be hard to discern whether our physical
vacuum is the actual ground state or whether our laboratory would be standing over a replica, as the differences are smaller than 
the model's systematic uncertainties. 

\begin{figure}[h]
    \centering
    \includegraphics[width=\columnwidth]{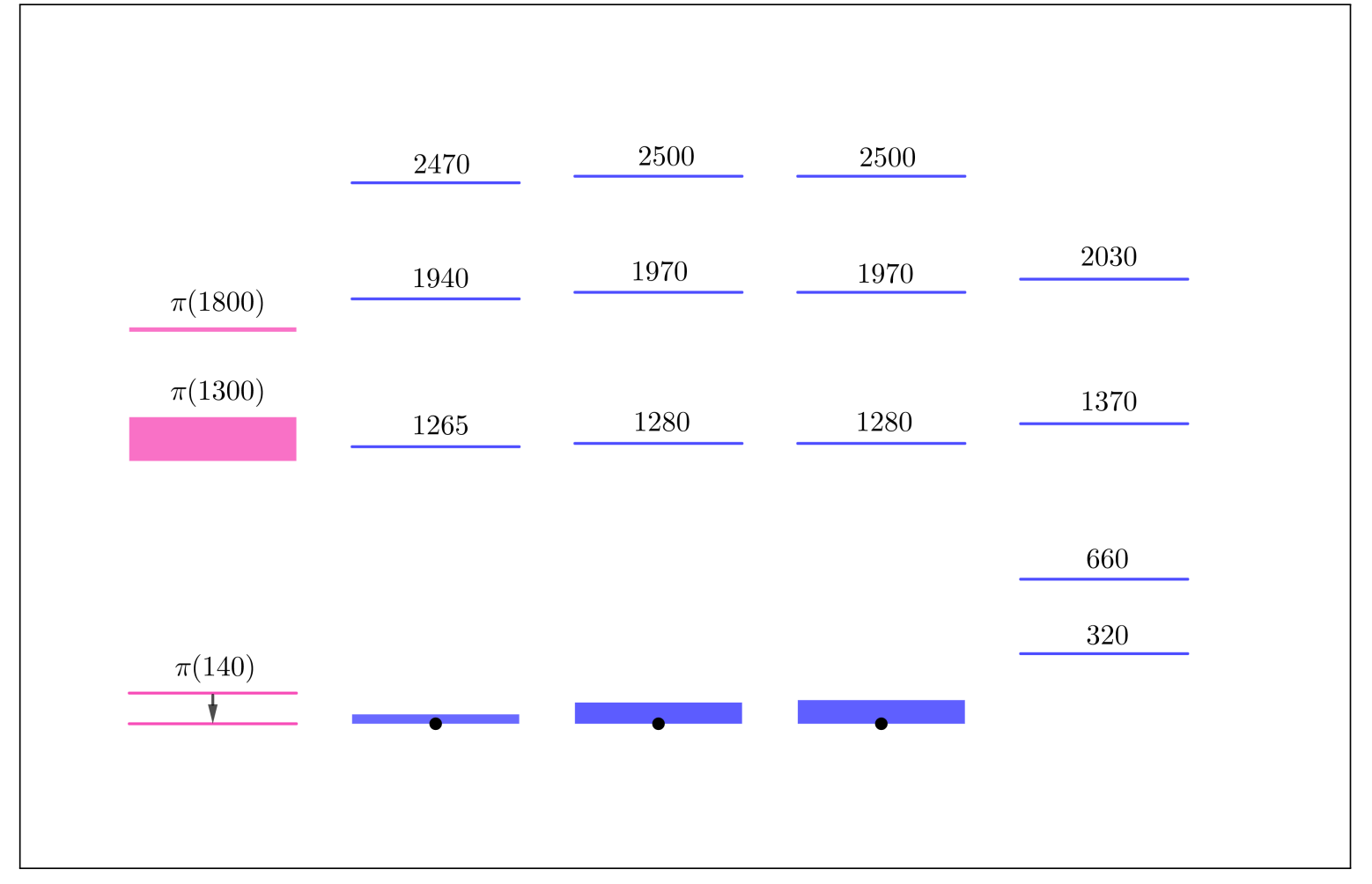}
    \caption{The pseudoscalar spectrum (including the pion) in the field theory employing the Cornell model (lines, blue online, from right to left the underlying vacuum is $|0\rangle$, $|\Omega''\rangle$, $|\Omega'\rangle$, $|\Omega\rangle$), compared with the experimental ones (leftmost, magenta online) and those from the harmonic oscillator potential ground states (round dots).  }
    \label{fig:pions}
\end{figure}

The spectrum is depicted in figure~\ref{fig:pions}.

We now switch the quark mass on, to either $m_q$=1 MeV or $m_q$=5 MeV. In this case we only have one replica, as already discussed. 
These masses are insignificant to make a difference for the excitations above a GeV, so we calculate only the first, ground-state $0^{-+}$ pion. Table~\ref{tabmq} gives this reduced spectrum.
\begin{table}[]
\centering
\caption{Dependence of the pion mass with the quark mass and the choice of vacuum function $\phi_i$ (compare with the experimental isospin-averaged mass $m_\pi=138$ MeV).}
 \label{tabmq}
\begin{tabular}{|c|c|c|c|} \hline
\textbf{Vacuum} & $m_q$=0 & $m_q$=1 & $m_q$=5  \\
\hline
$|\Omega\rangle$  & 0 & 150 & 390   \\ \hline
$|\Omega'\rangle$ & 0 & 468 &   751     \\ \hline
\end{tabular}
\end{table}

We can see that the ground--state BCS $|\Omega\rangle$ vacuum, with $m_q=1$ MeV, yields a satisfactory mass, although for too small a current mass. The Gell-Mann-Oakes-Renner relation~\cite{Gell-Mann:1968hlm,Cahill:1996bv} guarantees that any small pion mass can be reproduced by a small enough quark mass, 
and the RPA provides a dynamical implementation of the theorem, so the success here is unsurprising.  
The replica needs an even smaller quark mass to produce the physical pion. 

\begin{figure}[h]
    \centering
    \includegraphics[width=0.9\columnwidth]{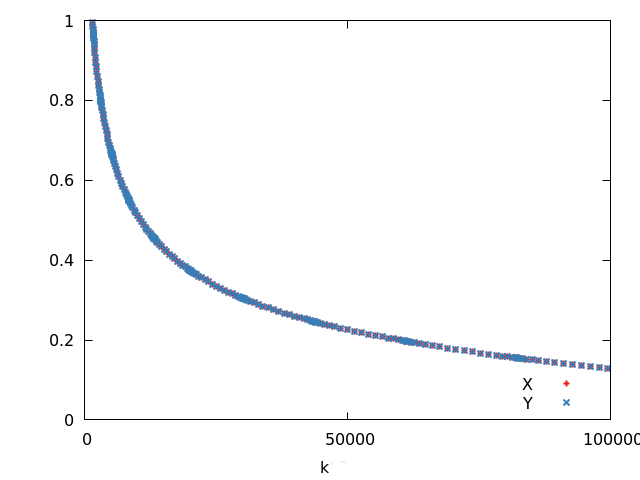}
    \caption{Radial wavefunctions $X$ and $Y$ as functions of the momentum $k$. The two sets of symbols fall on the same curve showing $Y=X$ for $m_q=0$.}
    \label{fig:XYequal}
\end{figure}

If the Tamm-Dancoff (or simply Schr\"odinger) approximation is adopted, in which the backpropagating component $Y$ is neglected, 
then the GMOR relation fails in the model and we obtain pion masses (in MeV) of 531 and 576 for $m_q$=1 and $m_q$=5, respectively, over the BCS ground state.

That is a poor approximation for the pion because in the chiral limit, $Y=X$ as shown in figure~(\ref{fig:XYequal}).

We also quickly discuss the vector mesons built over all these background states. The best known of those
is the $\rho(770)$ whose mass comes reasonably close to the experimental one, for $m_q=0$, for all the replicae and
even the vacuum of the free Hamiltonian! (see table~\ref{table:Erho}).
We will not spell here the somewhat laborious RPA equations for the coupled $s$- and $d$-wave channels which can be found in 
our previous work~\cite{Bicudo:1989si,Llanes-Estrada:2004edu}.

\begin{table}[h]
	\centering
 	\caption{Energy eigenvalues, in MeV, for vector mesons ($\rho (770)$ and $s$, $d$ wave excitations)
	over the perturbative vacuum $\ket{0}$, the BCS one $\ket{\Omega}$, and its two replicae.
	(Rounded off to 1 MeV for the ground state meson, 5 MeV for its first excitation, and 10 MeV for the rest.)}
	\begin{tabular}{c|ccccc}
		\hline \hline
		  \textbf{Vacuum}& ${\rho(770)}$& $\rho(1450)$ & $\rho(1570)$ & $\rho(1700)$ & $\rho(1900)$ \\ \hline 
     	$\boldsymbol{\ket{\Omega}}$ &749  & 1325 & 1500 & 1940 & 2090\\
        $\boldsymbol{\ket{\Omega'}}$ &796  & 1195 & 1410 &  1730 & 1940\\
        $\boldsymbol{\ket{\Omega''}}$& 811  & 1195 & 1440 & 1740& 1990 \\
        $\boldsymbol{\ket{0}}$& 812 & 1195 & 1440 & 1740 & 1990 \\
       \textbf{Experim.}\tablefootnote{Data from the \textit{Particle Data Group} compilation~\cite{ParticleDataGroup:2018ovx}.}& 775$\pm$($<$1) & 1465$\pm$25 & 1570$\pm$62 & 1720$\pm$20 & 1909$\pm$32 \\  \hline \hline
	\end{tabular}
    \label{table:Erho}
\end{table}%
As the table shows, the vector mesons constructed over the various $\phi_i$ vacua have masses with overall similar agreement to the experiment, so it would be difficult to distinguish over which vacuum our experiments reside from the meson spectrum alone. 

\begin{figure}
    \centering
    \includegraphics[width=\columnwidth]{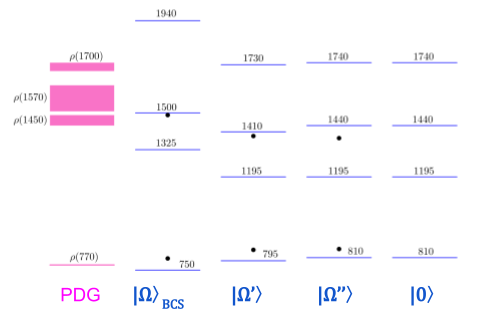}
    \caption{The vector spectrum in the field theory employing the Cornell model (lines, blue online) over the various reference vacuum states, compared with the experimental ones and those from the harmonic oscillator potential (round dots). }
    \label{fig:vectors}
\end{figure}

We again plot a Grotrian diagram with this spectrum in figure~\ref{fig:vectors}.

To complete this section, we note that a cursory variational computation was performed to obtain nucleon masses with the methods of~\cite{Bicudo:2009cr}. We obtained mass differences $M_N(\Omega')-M_N(\Omega)\sim 100$ MeV and $M_N(\Omega'')-M_N(\Omega')\sim 300$ MeV, albeit with large errors due to the Monte Carlo computation of the three-body matrix elements, so all we dare state is that the decrease of the constituent quark mass over the replicae pushes the baryon masses to somewhat lighter values as compared with the ground-state BCS vacuum.

%%%%%%%%%%%%%%%%%%%%%%%%%%%%%%%%%%%%%%%%%%%%%%%%%%%%%%%%%%%%%%%%%%%%%%%%
\section{Vacuum condensate and decay constant}\label{sec:condensateandfpi}
%%%%%%%%%%%%%%%%%%%%%%%%%%%%%%%%%%%%%%%%%%%%%%%%%%%%%%%%%%%%%%%%%%%%%%%%

Once the mass gap has been solved, we can calculate the vacuum quark-antiquark condensate given by 
\begin{equation}
    \langle \Bar{q} q \rangle \equiv \langle \Omega| \Bar{\Psi}(0) \Psi(0)|\Omega \rangle=-\dfrac{3}{\pi^2}\int_0^\infty \dif k \, k^2\sin \phi_k.
\end{equation}
This condensate is quadratically UV divergent for finite $m_q\neq0$, so  beyond the chiral limit we need to regulate this by subtracting the trivial vacuum contribution
\begin{equation}
\label{eq:conden}
    \langle \Bar{q} q \rangle_{reg}=-\dfrac{3}{\pi^2}\int_0^\infty \dif k \,k^2 \left(\sin \phi_k-\dfrac{m_q}{E_k}\right)
\end{equation}
and measure the condensate with respect to that zero point.

We have numerically calculated the quark condensate for each solution $\phi_i$ by means of Eq.~(\ref{eq:conden}). In the chiral limit $m_q=0$ there is no need for the subtraction and can also compute over the second replica: we have
\begin{align}
&\langle \Bar{q}q\rangle_{\phi^0}=-(178 \text{ MeV})^3,\\
\label{eq:c1}
&\langle \Bar{q}q\rangle_{\phi^1}=(73 \text{ MeV})^3,\\
\label{eq:c2}
&\langle \Bar{q}q\rangle_{\phi^2}=-(61 \text{ MeV})^3, \\
&\langle \Bar{q}q\rangle_{\phi^{\rm pert}}\equiv 0\ . \label{eq:cbcs}
\end{align}
Turning off the hyperfine interaction of Eq.~(\ref{eq:hyper}), the value of the condensate in the BCS ground state decreases in absolute value to about $-(120 \text{ MeV})^3$.
In turn, adding a small quark mass $m_q=1 \text{ MeV}$ we find, after mass subtraction, the increased values
\begin{align}
\label{eq:cbcss}
&\langle \Bar{q}q\rangle_{\phi^0}=-(189 \text{ MeV})^3,\\
\label{eq:c11}
&\langle \Bar{q}q\rangle_{\phi^1}=-(111 \text{ MeV})^3.
\end{align}
First, we can compare the quark condensate in the BCS solution $\phi_k^0$ with recent lattice estimations of this value~\cite{Gubler:2018ctz}. Our result is smaller than the latest lattice calculations of $\langle \Bar{q}q\rangle_{\phi^0}=-(272 \text{ MeV})^3$.  We therefore conclude an improved model may be needed to reach the lattice estimation for the condensate, including higher order terms in the kernel of $V_C$. This is unsurprising for a model approach. 

We have also recovered the small quark mass because it will help us understand the sign of the quark condensate. 
 First, consider the constituent masses $M(k)$ we have shown in Fig.~\ref{fig:masas} and \ref{fig:masas2}. We can see in these figures the dressed mass has negative parts, seemingly unphysical. However, $M(k)$ has to be understood as an auxiliary function for a confined quark, not a physical mass. In fact, the masses of the physical mesons have been calculated over the BCS vacuum and over the replicae and are positive as already seen in Sec.~\ref{sec:mesons}. Now consider the quark condensate calculated in Eq.~(\ref{eq:conden}). We can see it has an explicit minus sign and the result is expected to be negative, which is what happens in all the solutions except for the first replica in the chiral limit, as patent in Eq.~(\ref{eq:c1}). We can understand the change in the sign making use of the Gell-Mann-Oakes-Renner relation
\begin{equation}
    -\langle\Bar{q}q\rangle m_q(\mu)=m^2_\pi f^2_\pi
\end{equation}
where $m_\pi$ and $f^2_\pi$ are respectively the mass and decay constant of the pion and $\mu$ is the renormalization scale. This relation yields several interesting results. From it, one understands that what is a physical quantity is not the condensate or the quark mass, but the product of them (eventually, $m_q(\mu)\rightarrow M(k=\mu)$). In that case, the left side of the identity has to be positive, so if we have a positive condensate, the dressed mass has to be negative. And this is exactly what happens for the first replica in the chiral limit, as we can see in Fig.~\ref{fig:masas}, the dressed mass has mainly negative values. Notice the opposite happens with $m_q=1 \text{ MeV}$, the quark condensate is negative (Eq.~(\ref{eq:c11})) because the dressed quark mass is positive (Fig.~\ref{fig:masas2}).

We now proceed to discuss the pion decay constant (see {\it e.g.}~\cite{Gasser:2010wz,Narison:2015nxh} for further discussion on such decay constants). Although the eigenvalue problem is independent of the wavefunction normalization, we need to work it out to compute matrix elements.  Giving the pion operators a canonical behaviour
yields the normalization of the wavefunctions in Eq.~(\ref{RPApions}), 
\begin{eqnarray}
 \langle {\rm RPA}| Q_{\rm RPA}Q_{\rm RPA}^\dagger-Q_{\rm RPA}^\dagger Q_{\rm RPA})|{\rm RPA}\rangle  &=& \\ \nonumber
   \langle \big[\pi^a,\pi^{b \dagger}\big]\rangle &=&(2\pi)^3\delta^3(\vec{p}-\vec{q})
\end{eqnarray}
that yields an expression to be normalized, with the pion at rest, 
\begin{eqnarray}
    \sum_{\lambda_{1,2}\mu_{1,2}i_{1,2}j_{1,2}}\int\int\frac{d^3k_1d^3k_2}{(2\pi)^6}\frac{1}{4\pi}\frac{(\sigma_2)_{\lambda_1\mu_1}(\sigma_2)_{\lambda_2\mu_2}\delta_{i_1j_1}\delta_{i_2j_2}}{6} \nonumber \\
    \big[X^{\nu *}(k_1)X^\nu(k_2)(2\pi)^6\delta_{12}-Y^{\nu *}(k_1)Y^\nu(k_2)(2\pi)^6\delta_{12}\big] \nonumber \\
\end{eqnarray}
that finally results in an expression, in terms of a certain $N$ factor that collects the various contributions, 
\begin{equation}\label{normRPA}
    \int^\infty_0 dk k^2 \big(|X^\nu(k)|^2-|Y^\nu(k)|^2\big)= (2\pi)^3N^2 \ .
\end{equation}
Normalizing now $X\to X/N$, $Y\to Y/N$, the canonical commutation relation is satisfied. 
With this we are ready to compute the decay constant,

Lorentz invariance dictates the following parametrization of the matrix element of the axial current,
controlling the pion's weak decay,
\begin{equation}\label{deffpi}
    \langle\Omega|A_\mu (0)|P(\vec{p})\rangle=\frac{1}{\sqrt{E_p}}f_\pi p_\mu \ .
\end{equation}

The pion decay constant is therein $f_\pi$, $p_\mu$  the pion four-momentum, $E_p$ its relativistic energy and
 $A_\mu (\vec{x})=\overline{\Psi}(\vec{x})\gamma_\mu \gamma_5\Psi(\vec{x})$. The associated chiral charge is
\begin{equation}
    Q_5=\int d\vec{x}A_0(\vec{x})=\int d\vec{x}\Psi^\dagger(\vec{x})\gamma_5\Psi(\vec{x})
\end{equation}
and for a pion at rest, Eq.~(\ref{deffpi}) becomes
\begin{equation}
    f_\pi=\frac{1}{\sqrt{M_P}}\langle \Omega |\Psi^\dagger(0)\gamma_5\Psi(0)|\pi^\dagger(0)\rangle\ .
\end{equation}

A straightforward computation yields
\begin{eqnarray}
     \langle\Omega_{RPA}|\big[\Psi^\dagger(0)\gamma_5\Psi(0), Q^\dagger(RPA)\big]|\Omega_{RPA}\rangle
     \nonumber \\ =\int^\infty_0 dp \frac{\sqrt{24\pi}}{(2\pi)^3} \ s_p \ \big[ X^\nu(p)-Y^\nu(p)\big]
\end{eqnarray}
and we can finally read off the pion decay constant in terms of the two pion wavefunctions
\begin{equation}
    f_\pi^{RPA}=\frac{1}{\pi\sqrt{(2\pi)^3}\sqrt{M_\pi}}\int^\infty_0 dk \ s_k\big( X^\nu(k)-Y^\nu(k)\big)\ .
\end{equation}
Near the chiral limit, when $Y\simeq X$ implements a massless pion as an exact Goldstone boson, 
the zero in this expression is necessary to compensate the zero in Eq.~(\ref{normRPA}), that here appears in 
the denominator normalizing $X$ and $Y$. The result, although numerically delicate, is finite in the chiral limit.

Table~\ref{fig:tabfpi} shows the numerical values obtained over the various vacua, where the normalization is such that the physical experimental value would correspond to $f_\pi=93$ MeV.

\begin{table}[h]
\centering
\caption{Dependence of the computed decay constant with the vacuum chosen and the quark mass.}
 \label{fig:tabfpi}
\begin{tabular}{|c|c|c|c|} \hline
\textbf{Vacuum} & $m_q$=0 & $m_q$=1 & $m_q$=5  \\
\hline
$|\Omega\rangle$  & 21 & 23    & 30      \\ \hline
$|\Omega'\rangle$ & 1 &   2   & 2       \\ \hline
$|\Omega''\rangle$  & 0.08 &  --   &    -- \\ \hline
$|0\rangle $ & $\simeq$ 0 &    --    &      --   \\ \hline
\end{tabular}
\end{table}

The apparent ratio of the physical value to the computed one over the BCS vacuum is a large error factor of order 3-4, as has been
known~\cite{Llanes-Estrada:2001bgw} (remember that an error $O(\varepsilon^2)$ in an eigenvalue entails a larger error $O(\varepsilon)$ in the eigenvector,  and thus, in its transition matrix elements).  
Still, the order of magnitude is at least right if computing on this ground state BCS vacuum, 
whereas the replicae provide even smaller values of $f_\pi$, now off by orders of magnitude: it is recomforting to find that the ground-state BCS state is the one with the correct units at least.

\begin{figure}
    \centering
    \includegraphics[width=0.9\columnwidth]{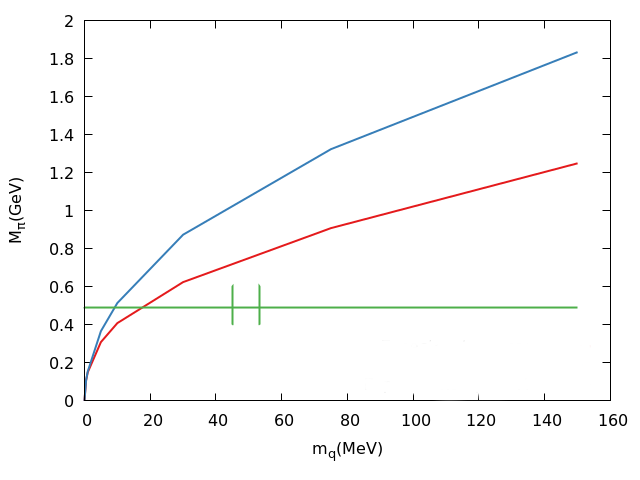}
    \caption{ The Gell-Mann-Oakes-Renner relation for the pseudoscalar mesons over the ground-state BCS vacuum. The 
    curves ascend too quickly. (The lower one is obtained from the calculated mass spectrum $M(m)$ and the upper one from 
    the GMOR relation from our computed $f_\pi$ and regularized condensate, they converge at low quark mass).
The green marks show the experimental value of the Kaon $K^-$ mass and the flavor--averaged quark mass interval $m_q\in (90, 107)$ MeV,
obtained from tabulated strange and light quark masses at a scale of 2 GeV. Ideally, the curves would have passed through that interval,
but $f_\pi$ in the model is too small.}
    \label{fig:GMOR}
\end{figure}

To further explore the interplay of the various pion observables, we can plot the GMOR relation
\begin{equation}
    M_\pi^2=\left(-\frac{2m_q\langle \overline{qq}\rangle}{f^2_\pi}\right) 
\end{equation}
that sets the constant of the $M_\pi\propto \sqrt{m_q}$ proportionality. This is done in figure~\ref{fig:GMOR}.
\\

While the model's pion mass has the correct functional dependence on the quark mass $M_\pi\propto\sqrt{m_q}$ as demonstrated 
early on~\cite{Llanes-Estrada:1999nat}, the curves ascend too quickly. 
The first line has meson masses obtained via the matrix element of the axial current, proceeding in the direction
 $f_\pi \to M_\pi$, in the GMOR formula. 
  The second line comes from the pion mass calculated explicitly at various quark masses. Both curves have similar shapes and diverge away from each other
  for larger $m_q$.

Comparing with the experimental data (the position of the 497 MeV kaon mass at lattice-extracted quark masses is marked in the figure, and lies to the right of the ascending curves), we observe a model difference among the pseudoscalar mass values (with fixed $m_q$) either directly calculated or extracted from $f_\pi$ and the condensate, in consonance with the low values for $f_\pi$ that the model yields.
Indeed, the kaon takes its correct mass in the model at quark masses that are at least a factor of 2 too small when compared with lattice gauge theory extractions. This is fine in a model, the quark mass is a fitting parameter: but it would be interesting to know why the decay constant comes out too small.

A possible reason is that the instantaneous potential has a scale fixed by  $m_g$, the gluon scale. 
An important result by Zwanziger~\cite{Zwanziger:2002sh} is that the Coulomb potential grows with $r$ faster than the static
quark-antiquark potential as measured in the lattice (or the bottomonium experimental spectrum), because the static potential in
Coulomb gauge includes the Coulomb potential and the average contribution of physical transverse gluons that lowers the original Coulomb string tension.

That is, the color potential that yields the spectrum at large distances where the linear tail is measurable
is less intense than the fundamental potential in the Hamiltonian. Thus, increasing the string tension or equivalently $m_g$
raises $f_\pi$ (but then the potential should be partially screened through some dynamical mechanism to recreate the spectrum).
We see no point in working this scenario out in model terms and remain content to have a good spectrum at the price of too small a decay constant, while the decay constant over the replicae comes out much smaller.

%%%%%%%%%%%%%%%%%%%%%%%%%%%%%%%%%%%%%%%%%%%%%%%%%%%%%%%%%%%%%%%%%%%%%%%%
\section{How to search for replicae?} \label{sec:latticesearch}
%%%%%%%%%%%%%%%%%%%%%%%%%%%%%%%%%%%%%%%%%%%%%%%%%%%%%%%%%%%%%%%%%%%%%%%%

%%%%%%%%%%%%%%%%%%%%%%%%%%%
\subsection{In Heavy Ion Collisions}
%%%%%%%%%%%%%%%%%%%%%%%%%%%
Taking as zero of the energy density the BCS ground state, we found for the first and second replicae the following energy 
densities,
\begin{align}
    &\rho_1=(0.1181 m_g)^4  =3.30 \text{ MeV/}\text{fm}^3,\\
    &\rho_2=(0.1182 m_g)^4= 3.31 \text{ MeV/}\text{fm}^3,
\end{align}
where $m_g=600 \text{ MeV}$ is the scale for the theory and they are measured from the BCS ground state $\phi_k^0$.

We can then calculate at what temperature a gas of pions (that are the first excitation appearing over $|\Omega_i\rangle$) can populate the excited vacua. This will give us some insight in how much temperature would be necessary to reach the replicas and if the needed temperature is lower than the phase transition from gas to plasma, allowing the gas to occupy the replicas.

\begin{figure}[h]
\centering
\includegraphics[scale=0.23]{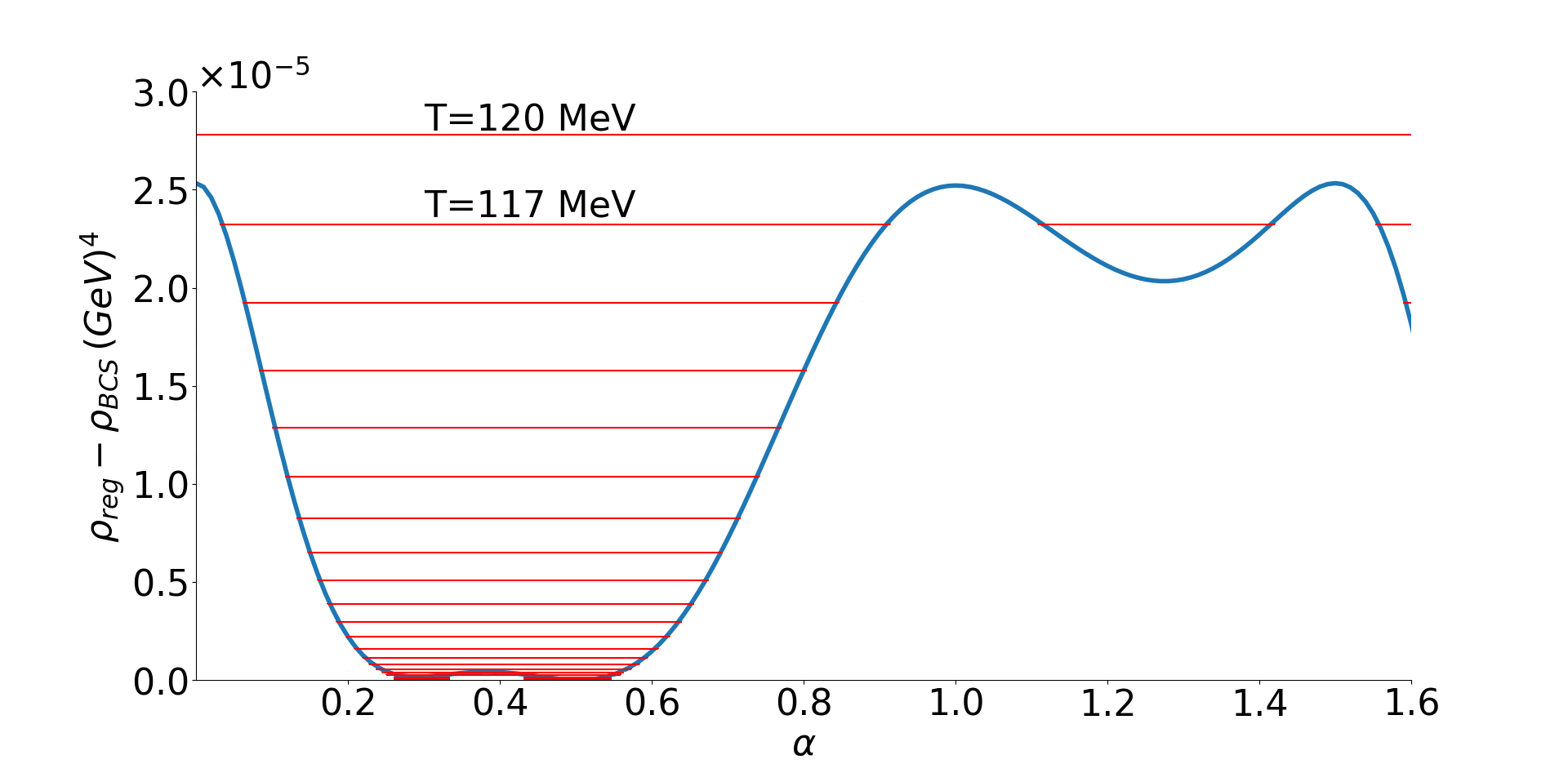}
\caption{\label{fig:piongas} Energy-density levels of a gas of pions superposed over $\langle H\rangle(\alpha)$ for the parametrization of 
Eq.~(\ref{interpol2infull}).
 From top to bottom, each horizontal red line increases the temperature by $3 \text{ MeV}$.}
\end{figure}

We use the energy density of the gas of pions calculated in the frame of reference where the gas is at rest (see \cite{GomezFonfria:2021lbu} and references therein)
\begin{equation}
  \rho_{\text{gas}}(T)=g\int \dfrac{\dif^3 k}{(2\pi)^3}\dfrac{\sqrt{m^2_\pi+k^2}}{e^{\sqrt{m^2_\pi+k^2}/T}-1}
\end{equation}
where $g=3$ is the degeneration for the pion isospin triplet, $m_\pi$ is the pion mass and $T$ is the temperature. We can easily calculate this integral numerically for a given temperature and then represent this energy density as energy levels (horizontal lines)
superposed over the replica landscape of Fig.~\ref{fig:birdseyeinterpol}. Then we can just read off at what temperature $T$ the gas is energetic enough to populate the replicae, and this is shown in Fig.~\ref{fig:piongas}. 
We can see both replicae can be occupied for temperatures near $T=117-120 \text{ MeV}$. This result is compatible with the gas jumping to the replicae before reaching the quark-gluon plasma phase, since the transition is of order $T_c\approx 170 \text{ MeV}$. Moreover, it shows these replicae could have been populated in the early stages of the universe, and then perhaps   decay with the decrease of temperature to the BCS vacuum.

Because the replicated vacua break the same chiral symmetry as the ground state, although less intensely so, we expect no first order phase transition in populating them, but a smooth crossover which is consistent with phenomenology of heavy ion collisions~\cite{Glozman:2022zpy} in approaching a chirally restored phase. Any replicae serve as stepping stones, as bubbles of less-intensely broken achiral phase can appear before full chiral symmetry restoration.

At large temperature,
a replica cannot duplicate the number of hadron degrees of freedom, as the entire spectrum over the replica substitutes the spectrum over the ground state with analogous hadrons over both. However, at lower temperatures near those 120 MeV, the energy density needed to change the collision medium to the replica vacuum is detracted from the rest of the hadron gas: if produced, the replica acts as an energy reservoir. As the system expands and cools, that energy appears relatively late in the evolution and is directly fed to pions (not to heavy resonances which play no role in chiral symmetry breaking). One could for example look for individual collision events in which the pion temperature recorded from their momentum spectrum would be higher than that of heavier species with, for example, strangeness or maybe charm.  This behavior is opposite to the usual situation in which the heavier degrees of freedom, which decouple first at higher temperatures, are hotter, and thus it becomes a telltale sign of the replica.

%%%%%%%%%%%%%%%%%%%%%%%%%%%%%%%
\subsection{In lattice gauge theory}
%%%%%%%%%%%%%%%%%%%%%%%%%%%%%%%
Before drawing strong conclusions however, one would like to find out whether these replicae carry over from Hamiltonian model of QCD to full QCD; this is not obvious because of the nonlinear nature of the theory. It is then interesting to search for them in a lattice gauge theory computation.

Such search can be based on the characteristic features of a replica which are, as shown above in figure~\ref{fig:overlap},
\begin{itemize}
\item They appear as a continuum of excited scalar states with energy $E=\rho V$ proportional to the lattice volume.
\item Any matrix elements with the ground state or conventional hadrons thereon are exponentially suppressed with that volume.
\item Therefore, eliminating the volume, the overlaps fall exponentially with the energy of the scalar excitation, $\Gamma_{\rm any}\propto e^{-E}$\ .
\end{itemize}

 A way to address them is to start from
the Euclidean correlator
\begin{equation}
    \mathcal{C}\equiv \sum_{t} C(t)=\sum_{t} \expval{O(t)\Bar{O}(0)}
\end{equation}
propagating during a time $t$ an operator $O$ with the quantum numbers of a scalar state (for example a four-pion state). Inserting the spectral decomposition of the time-evolution operator one has
\begin{equation}
\begin{split}
   C(t)= \expval{O(t)\Bar{O}(0)}&=\sum_k \bra{0}\hat{O}\ket{k}e^{-tE_k}\bra{k}\hat{O}^\dagger\ket{0}\\
    &\propto e^{-t E_H}(1+\mathcal{O}(e^{-t\Delta E})),
    \end{split}
\end{equation}
with $E_H$ the mass of the ground state hadron in that channel, and those of the excited states reachable via the subleading terms with $\Delta E$. 

Fitting the exponential tails allows to extract the ground state and eventually excited ones. Although such calculations are performed at finite volume, once the lattice is large enough to contain the hadron inside, finite size effects that affect the extracted hadron masses  exponentially diminish with further volume increases. 

\begin{figure}[h]
\centering
\includegraphics[width=0.42\textwidth]{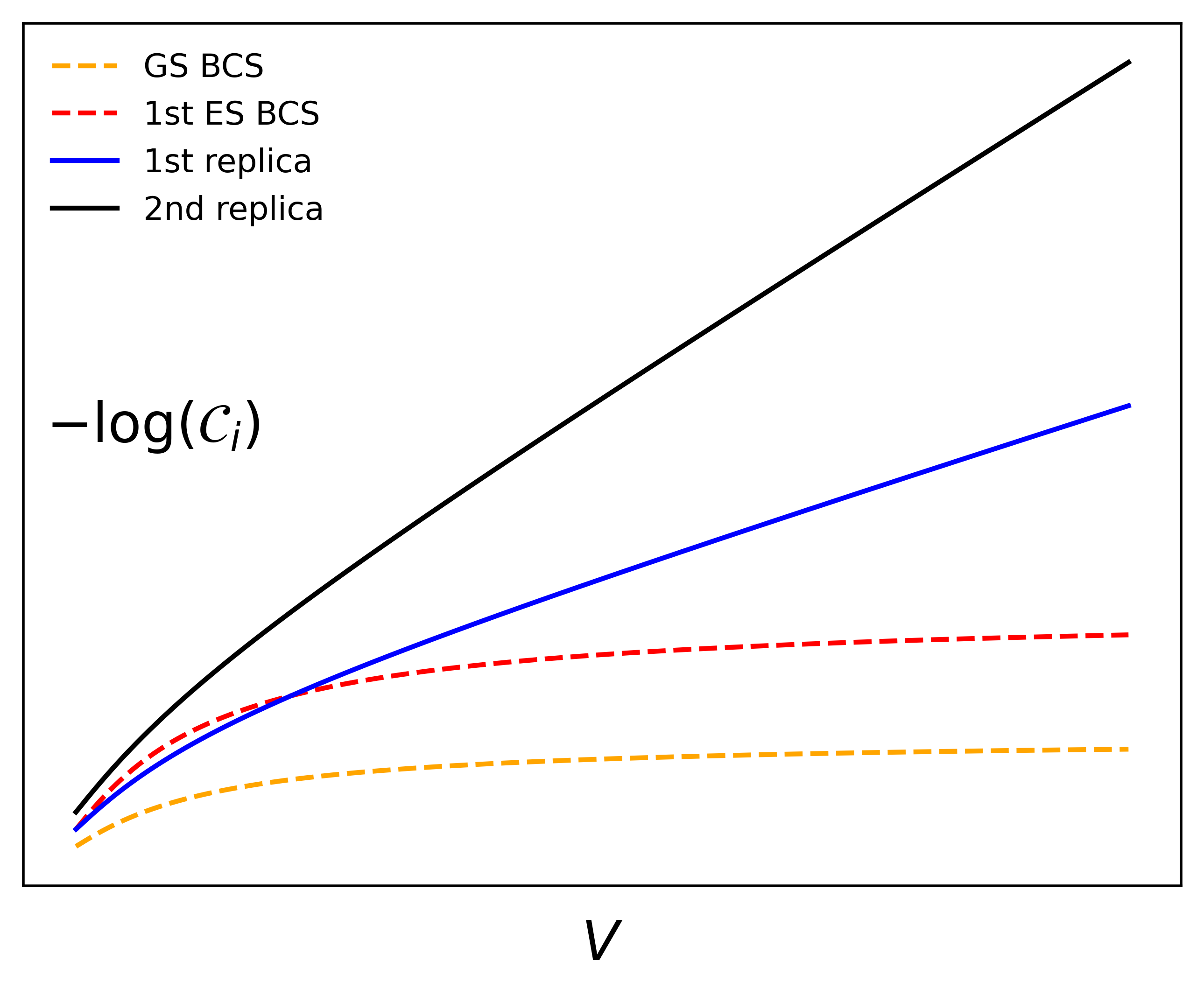}
\caption{\label{fig:latreplicae} Schematic view of the positively sloped linear behaviour (solid lines) with volume of the correlator exponents for the replicae, with a constant \emph{energy density} as opposed to the constant \emph{energy} plateau of ordinary hadrons built over the minimum-energy vacuum state (two shown, dashed lines).}
\end{figure}

However, the replicae if they are present in the theory would appear as scalar contributions with energy proportional to the volume. Hence, the part of the fully resummed correlator $\mathcal{C}$ which overlaps with the replicae would have a term whose exponent would scale linearly with the volume, while the states built over the ground-state vacuum would show an energy plateau after finite-size effects disappear (as the energy is fixed to be the mass of one hadron). 
That is, ordinary hadrons yield a decaying contribution to the correlator $C_H \propto  e^{-{\rm constant}\times t}$ whereas the replicae give a contribution
$C_R \propto e^{-{\rm constant} \times tV}$ whose exponent is proportional to both the time and the spatial volume: they are rather evanescent quantities. The distinction is illustrated in figure~\ref{fig:latreplicae}.

In this way, the existence of the replicae could be probed in an Euclidean lattice calculation. 

\newpage
%%%%%%%%%%%%%%%%%%%%%%%%%%%%%%
\section{Conclusions} \label{sec:conclusions}
%%%%%%%%%%%%%%%%%%%%%%%%%%%%%%

We have presented an updated in-depth study of the vacuum replicae of the chiral-symmetry breaking
vacuum in Hamiltonian models of Coulomb gauge-QCD. We have employed a previously well-studied model with relativistic spinors and a linear+Coulomb potential in momentum space, but the results are qualitatively the same as those obtained with a spherical harmonic oscillator as the confining potential, reported in earlier work~\cite{Bicudo:2019ryc}.
In particular, we obtain the same number of  replicae below the perturbative vacuum and extract spectra that behave in a similar way over each of the ground- and excited-state vacua.
We concur with earlier work in the fact that the replicae are stable to few-body excitations (all hadrons over each of the replicae have real, nonnegative masses).

Beyond a somewhat more realistic interaction, where we have advanced more significantly is in the understanding of the collective nature of the replicae in the function space of the chiral-symmetry breaking gap angle (the Bogoliubov angle $\theta_k$ or equivalently, the BCS mass gap angle $\phi_k$). Here we have seen that the replicae, as well as the perturbative vacuum, present negative-eigenvalue modes which make them unstable to collective excitations along those directions, with the overlaps falling exponentially with the quantization volume. 

The study of those eigenvalues suggests that our numerical exploration has found all the possible solutions of the BCS gap equation with negative energy density in the Coulomb gauge North Carolina  State model~\cite{Robertson:1998va}. 

We have also observed that, while the spectra of hadrons built over the ground state and over the replica are not too dissimilar (note that the perturbative vacuum is chirally symmetric, so over that one the spectrum does show different features), matrix elements which involve the gap angle $\phi(k)$ such as decay constants, condensates and others, allows to distinguish two ground states much better. 

Beyond our field-theory approach, we have shown how a future lattice study may be carried out to try to confirm that these replicae are a feature of full chromodynamics. Since the model field theory captures the chiral symmetry breaking aspects of the full field theory and the replicae appear in the spontaneous chiral-symmetry breaking equations, we expect them to be generic. It would be interesting to know whether the covariant Dyson-Schwinger approach~\cite{Roberts:2023lap,Alkofer:2023lrl,Fischer:2018sdj} to chiral-symmetry breaking, which also features a gap equation (that for the fermion propagator) presents more than one solution with realistic interactions, beyond the simple scalar model considered so far~\cite{Llanes-Estrada:2006bxu}.
Should this be accomplished, we believe that the replicae can leave observable signals in systems at large temperature such as relativistic heavy-ion collisions
or in the early universe. 
\\ $ \phantom{leave some space}$ \\

%%%%%%%%%%%%%%%%%%%%%%%%%%%%%%%%%%%%%%%%%%%%%%%%%%%%%%%%%%%%%%%%%%%%%%%%
\acknowledgments
%%%%%%%%%%%%%%%%%%%%%%%%%%%%%%%%%%%%%%%%%%%%%%%%%%%%%%%%%%%%%%%%%%%%%%%%

FJLE thanks Enrique S\'anchez Ib\'a\~nez for some assistance at early stages of the project.
Work partially supported by grant PID2022-137003NB-I00 financed by spanish
MCIN/AEI/10.13039/501100011033/ and FEDER programs, as well as further EU support under grant 824093 (STRONG2020); CeFEMA under the portuguese FCT contract for R\&D Units UIDB/04540/2020; and EGV thanks IPARCOS for partial financial support during the development of this work and is currently supported by german DFG (Collaborative Research Center CRC-TR 211 ``Strong-interaction matter under
extreme conditions'' - project number 315477589 - TRR 211). 
\\

%%%%%%%%%%%%%%%%%%%%%%%%%%%%%%%%%%%%%%%%%%%%%%%%%%%%%%%%%%%%%%%%%%%%%%%%
\emph{Disclosure of author responsibility.}
%%%%%%%%%%%%%%%%%%%%%%%%%%%%%%%%%%%%%%%%%%%%%%%%%%%%%%%%%%%%%%%%%%%%%%%%
P. Bicudo and J. Ribeiro have devised and guided the investigation.   
Section~\ref{sec:slices} relies largely on the work of E. Garnacho and J. Vallejo. 
The meson computations of section~\ref{sec:mesons} are due to V. Serrano and L.P. Molina.
Section~\ref{sec:condensateandfpi} has been worked out by E. Garnacho and L.P. Molina, 
whereas E. Garnacho and F. Llanes have researched section~\ref{sec:latticesearch}.
F. Llanes has directed the execution of the investigation and prepared the manuscript.

\newpage
%%%%%%%%%%%%%%%%%%%%%%%%%%%%%%%%%%%%%%%%%%%%%%%%%%%%%%%%%%%%%%%%%%%%%%%%

\end{document}